\documentclass[fleqn,usenatbib]{mnras}
\usepackage{amsmath}
\usepackage{graphicx}
\usepackage{amssymb,txfonts}
\usepackage{esvect}

\def\kte{kT_{\rm e}}
\def\ktbb{kT_{\rm bb}}
\def\grbcomp{{\sc grbcomp}}
\def\xspec{{\rm XSPEC}}

\def\liso{L_{\rm iso}}
\def\eiso{E_{\rm iso}}
\def\ep{E_{\rm p}}
\def\epi{E^{\rm i}_{\rm p}}
\def\epobs{E^{\rm p}_{\rm obs}}

\def\ejet{E_{\rm jet}}
\def\eobs{E_{\rm obs}}

\def\thetaobs{\theta_{\rm obs}}
\def\thetajet{\theta_{\rm jet}}

\def\rjet{R_{\rm jet}}
\def\r12{R_{\rm 12}}

\def\sw{\textit{Neil Gehrels Swift Observatory}~}

\def\tp{t_{\rm p}}

\def\tspa{t_{\rm spa}}
\def\tgrb{t_{\rm grb}}
\def\eobs{E_{\rm obs}}

\def\costhobs{\cos\theta_{\rm obs}}
\def\sinthobs{\sin\theta_{\rm obs}}
\def\costhjet{\cos\theta_{\rm jet}}
\def\costheta{\cos\theta}

\def\sinphi{\sin\phi}
\def\cosalpha{\cos\alpha}

\def\sintheta{\sin\theta}
\def\sinphi{\sin\phi}

\def\domegaeff{d\Omega^{\rm eff}}
\def\gmax{\Gamma_{\rm max}}
\def\gmin{\Gamma_{\rm min}}

\def\Df{D_{\rm f}}
\def\dl{D_{\rm L}}
\def\da{D_{\rm A}}
\def\du{{\rm d}u}
\def\dphi{{\rm d\phi}}

\def\Ijet{\rm I_{\rm jet}}

\def\tgrb{t_{\rm grb}}
\def\sigmap{\sigma_{\rm P}}

\title[Jet model of gamma-ray bursts]
{A numerical jet model for the prompt emission of gamma--ray bursts}
\author[R. Farinelli et al.]
{Ruben Farinelli, $^{1}$ \thanks{E-mail: ruben.farinelli@inaf.it} Rupal Basak, $^{1, 2}$ 
Lorenzo Amati,$^{1}$ Cristiano Guidorzi,$^{1,2,3}$
\newauthor
Filippo Frontera $^{2}$
\\
$^1$INAF -- Osservatorio di Astrofisica e Scienza dello Spazio di Bologna, Via P. Gobetti 101, I-40129 Bologna, Italy\\
$^2$Dipartimento di Fisica e Scienze della Terra, Universit\`a di Ferrara, Via Saragat 1, I-44122 Ferrara, Italy\\
$^3$INFN – Sezione di Ferrara, Via Saragat 1, I-44122 Ferrara, Italy
}

\begin{document}

\date{Accepted . Received 2020 ; in original form }

\pagerange{\pageref{firstpage}--\pageref{lastpage}} \pubyear{2020}

\maketitle

\label{firstpage}

\begin{abstract}
Gamma-ray bursts (GRBs) are known to be highly collimated events, and are mostly detectable when they are seen on-axis or very nearly on-axis. However, GRBs can be seen from off-axis angles, and the recent detection of a short GRB associated to a gravitational wave event has conclusively shown such a scenario. The observer viewing angle plays an important role in the observable spectral shape and the energetic of such events. We present a numerical model which is based on the single-pulse approximation with emission from a top-hat jet and 
has been developed to investigate the effects of the observer viewing angle. 
We assume a conical jet parametrized by a radius $\rjet$,  half-opening angle $\thetajet$,  a comoving-frame emissivity law 
and an observer viewing angle $\thetaobs$, and then study the effects for the conditions $\thetaobs<\thetajet$
and $\thetaobs >\thetajet$. 
We present results considering a smoothly broken power-law emissivity law in jet comoving frame, albeit the model implementation easily allows to consider
other emissivity laws.
We find that the relation $\epi \propto \eiso^{0.5}$ (Amati relation) is naturally obtained from pure relativistic
kinematic when $\Gamma \ga 10$ and $\thetaobs < \thetajet$; on the contrary, when $\thetaobs > \thetajet$ it results $\epi \propto \eiso^{0.25}$. Using data from literature for a class of well-know sub-energetic GRBs, we show that their position in the $\epi-\eiso$ plane is consistent with
event observed off-axis.
The presented model is developed as a module to be integrated in spectral fitting software package XSPEC and can be used by the scientific community.

\end{abstract}

\section{Introduction}
\label{intro}


Gamma-ray bursts (GRB) remain one of the most debated transient phenomena even after decades of research. It is fairly established that a GRB is a highly collimated event powered by a relativistic jet (\citealt{sari99, aloy00, zhang03}), launched during the catastrophic death of a massive star (\citealt{woosley93}) or a coalescence event between two neutron stars or a neutron star - black hole pair (\citealt{eichler89,li98, abbott17a}). 
The radiation process of the prompt emission phase remains at the center of the debate ever since their discovery. From theoretical considerations, synchrotron  emission in an ordered or random magnetic field seeems to be a good description of the broadband spectra (\citealt{meszaros93, meszaros94, katz94, ghisellini99}). But, based on the spectral index below the peak energy and the efficiency of gamma-ray photon production, the models involving synchrotron process face some criticism (\citealt{crider97, preece98, kaneko06}, though see \citealt{burgess19}). 
Several alternative models have been proposed like synchrotron self-Compton (\citealt{dermer00, nakar09}), inverse-Compton scattering (\citealt{lazzati00, duran12}), different flavours of photospheric models (\citealt{beloborodov11, lundman13, begue15, ahlgren19}), as well as hybrid models in which different processes can also evolve in terms of their dominance (\citealt{zhang18}). 
Usually the prompt GRB spectra are fitted with phenomenological
models, such as cut-off powerlaw or the widely used Band function \citep{band93}. An additional photospheric component has been
sometimes detected and modeled with a blackbody \citep{ryde10}.
In addition, there have been some recent developments to employ more physically motivated models for the spectral fitting (\citealt{burgess19}). 
The first physical model (\grbcomp\, also released for the XSPEC package) applied to the GRB spectral analysis has been developed by \cite{titarchuk2012} and later tested on a sample of time-resolved spectra by \cite{frontera2013}.

The spectral models, whether empirical or driven by a physical scenario, mostly have an implicit assumption that the radiation received by the observer comes directly on-axis from the GRB jet (\citealt{proga03}). Due to the relativistic effects the jet is highly collimated and the radiation is indeed beamed within a very narrow cone, which justifies the assumption. However, it is highly probable to see a GRB from an off-axis angle, not necessarily outside the cone, and in general it can potentially alter the observables from an axisymmetric case (e.g., \citealt{yamazaki03, kathirgamaraju18}). The observer viewing angle ($\thetaobs$) thus becomes an important parameter in the analysis of the observed light curves and spectra. 

Naturally, the relatively nearby GRBs are the most interesting cases of highly off-axis GRBs. For instance, \citet{salafia16} calculated that a sizable fraction, anywhere between 10\% to 80\% of nearby GRBs (redshift, $z<0.1$) are detectable from an off-axis angle outside the jet cone by the \sw \citep{gehrels2004}.The recent discovery of a short GRB associated to a gravitational wave signal (GW 170817) has conclusively shown that the GRB was seen from an off-axis angle \citep{abbott17a, abbott17b, margutti17}. 

Another example is GRB 150101B ($z=0.13$)  which may have been seen from an off-axis angle of $13^\circ$ (\citealt{troja18}) while having a half-opening angle close to $9^{\circ}$ (\citealt{fong16}).
The nearby low-luminosity long GRB sample is also interesting as being faint and nearby there is a good probability that they may fall in the category of GRBs seen off-axis (\citealt{fynbo04}). However, several studies based on radio observation as well as the luminosity function and local rate have suggested that they might be intrinsically fainter rather than being seen off-axis, and can even belong to a different population compared to the cosmological high luminosity GRB class (\citealt{soderberg04, soderberg06, pian06, liang07}). On the other hand, hydrodynamic simulations of GRB jets show that the jet break signature of the off-axis GRBs can be delayed by several weeks and may remain hidden in the data (\citealt{vaneerten10}). The lack of jet break signature in \textit{Swift} sample is indeed consistent, and thus provides room for the alternative scenario. 

Some of the low-luminosity GRBs are found to be outliers of the Amati relation (\citealt{amati02, amati06}), hereafter AR, which is a relation between the intrinsic peak energy of the $EF(E)$ spectrum ($\epi$) and the isotropically-equivalent total emitted energy ($\eiso$).
They are GRB 980425, GRB 031203, GRB 080517, GRB 100316D, GRB 171205A (\citealt{campana06, ghisellini06, amati07, liang07, starling11, heussaff13, stanway15, delia18}). 

Several studies have been performed to investigate the observational effects of an off-axis jet (e.g., \citealt{yamazaki03, guidorzi2009,kathirgamaraju18}), albeit the
principally focused on the light curve shape and temporal variability.

The main goal of this work is to provide to the scientific community the first relativistic jet model for the X-ray spectral fitting package \xspec\ \citep{arnaud96}.
The intrinsic complexity of the jet physics \citep{meszaros01, ruiz02, zhang03, zhang04, morsony07, mizuta11} and the 
need to achieve a trade-off between computational speed
and model accuracy leads to introduce unavoidable simplifications in code development.
We have built a model based on the so-called
Single Pulse Approximation (SPA), where a top-hat relativistic jet instantaneously emits a flash
of radiation in the star frame \citep{yamazaki03}.

The jet can be viewed at any angle and 
we obtain the total spectra by
integrating over the different areas 
on the emitting surface. 
We allow to consider different emissivity laws in the comoving-frame.
This is thus not a radiative-transfer model, but it parametrises
the geometry of the emission in terms of jet radius
and opening angle as well as observer's viewing angle.

In addition to a simple top-hat jet, various theoretical arguments and hydrodynamic simulations,  as well as observations, have proposed GRB jets with angular structure (\citealt{berger03, basak15, margutti17, beniamini2019, salafia2020}), i.e. non-constant $\Gamma$-factor.
We then also consider the possibility of a structured jet 
and present some results, albeit in the release for
the \xspec\ package  we implemented the case of
a constant $\Gamma$-factor, to avoid having a too-high
number of free-parameters.

The paper is structured as follows: in Section \ref{model_formulation} we present the mathematical formulation of the model. In Section \ref{result} we report
results as a function of the input parameters and compare the case of constant and variable $\Gamma$ factor.
In Section \ref{amati_relation} we show the important
consequences at observational level between on-axis and off-axis cases in terms of the well-know $\epi-\liso$ or
$\epi-\eiso$ relations. The Discussion and Conclusions
are presented in Sections \ref{discussion} and \ref{conclusions}, respectively.

\begin{figure*}
  \hspace{0.1in}
\includegraphics[scale=0.25]{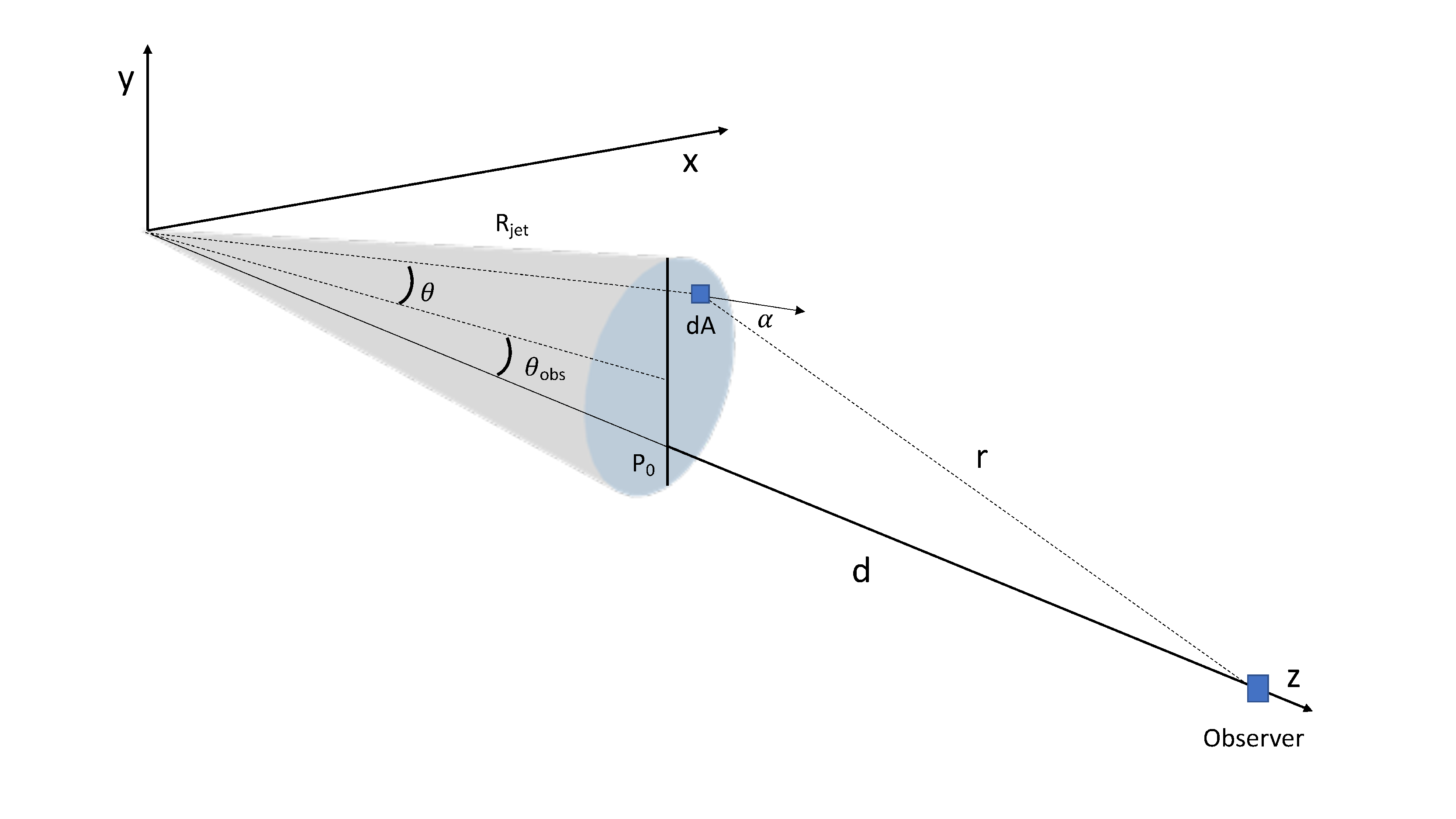}
\includegraphics[scale=0.25]{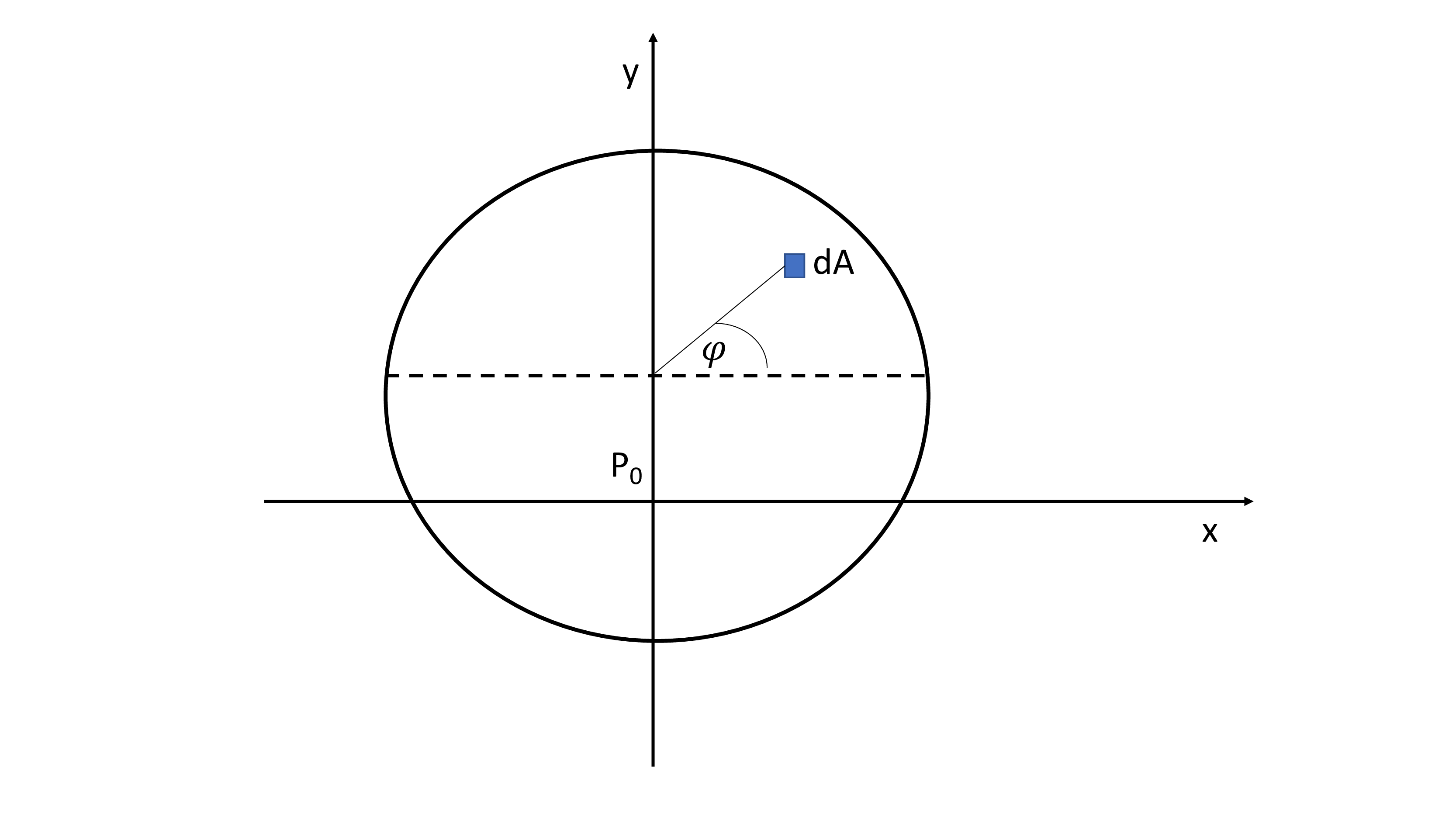}
 \includegraphics[scale=0.25]{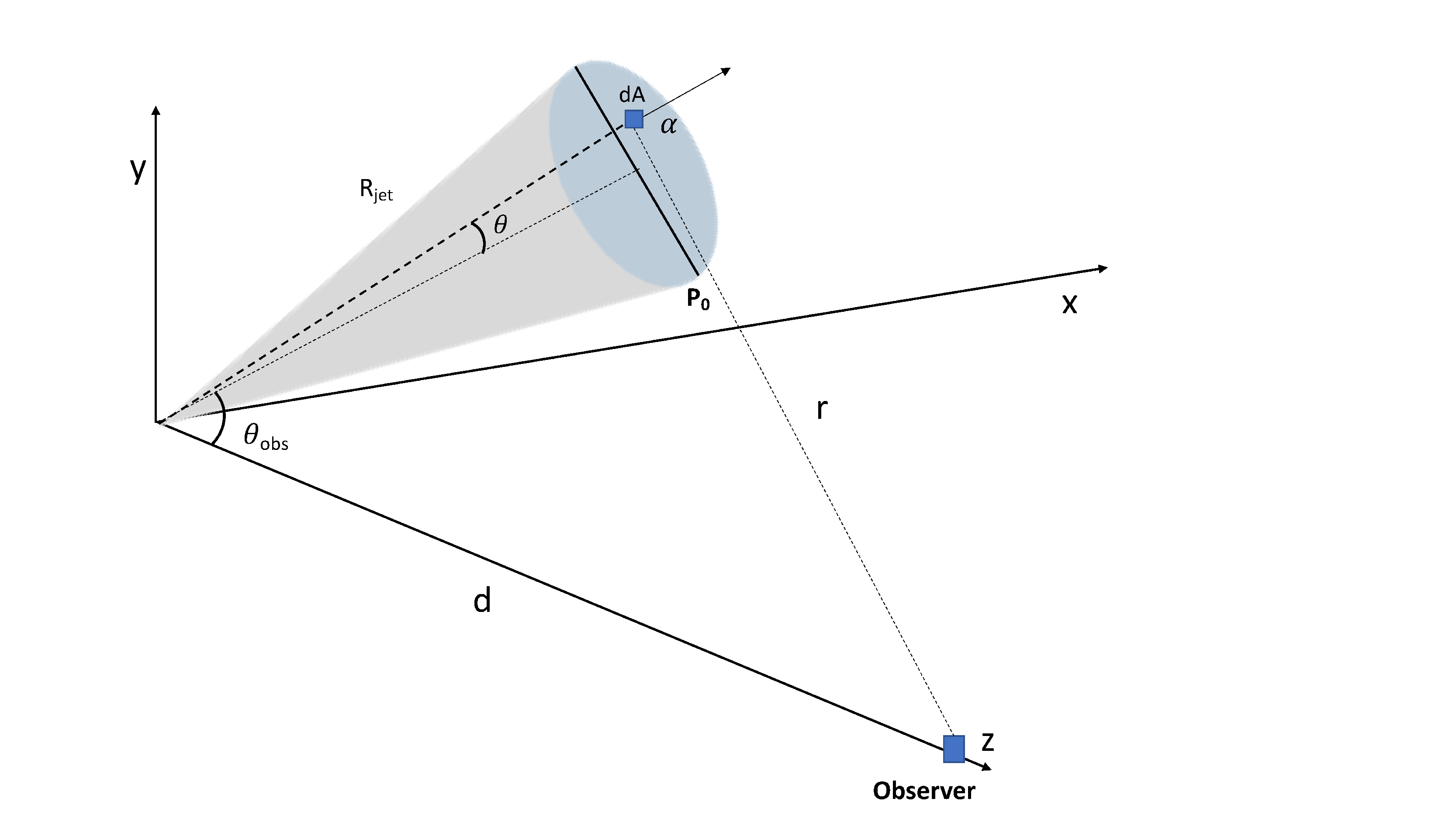}
 \includegraphics[scale=0.25]{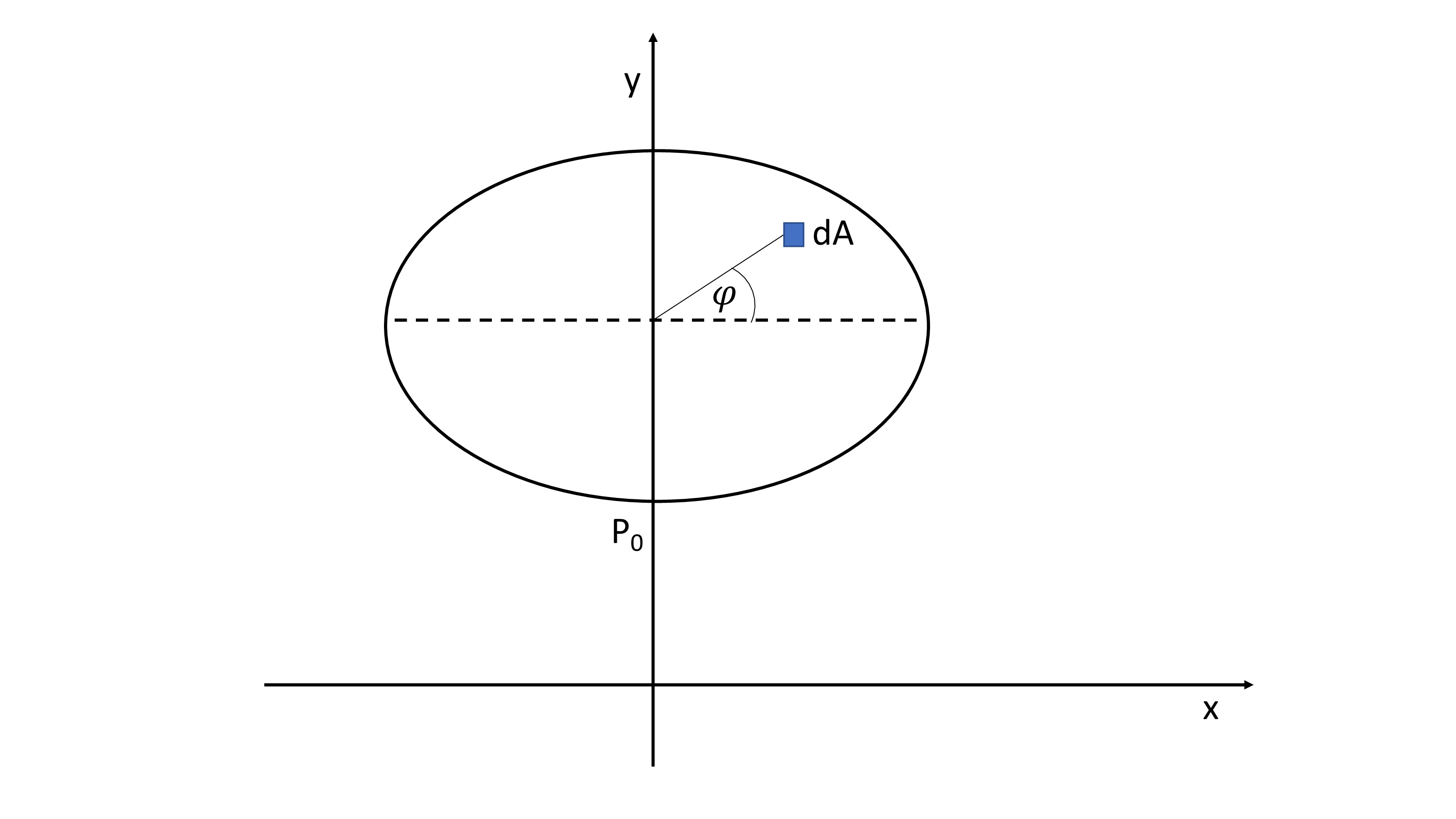}
\caption{Schematic view of the jet geometry for the on-axis (\emph{top-figures}) and off-axis (\emph{bottom-figures}) cases in a reference system $xyz$ where the $z$-axis is aligned towards the observer. For computing the spectrum (see equation [\ref{angular_integral_flux}]) it is more convenient to define the polar and azimuthal angles $\theta$ and $\phi$ in a reference system $XYZ$ rotated by an angle $\thetaobs$ around the $x$-axis and where the $Z$-axis is aligned with the jet axis. The point $P_0$ represents the region on the jet surface from where photons first reach the observer.}
\label{jet_figure}
\end{figure*}

\section{Geometric and physical definition of the jet model}
\label{model_formulation}

\subsection{Emission spectra}

We present here the mathematical details on which the jet geometry and emission are based. The most important parameters for the problem formulation are shown in Fig. \ref{jet_figure}. 
Let us first consider a reference system $XYZ$ where the  jet axis is aligned with the Z-axis. Applying a rotation of an angle $\thetaobs$ around the X-axis, the top-hat  cartesian coordinates in a system $xyz$ where the z-axis is directed towards the observer are

\begin{equation}
    \begin{array}{l} 
x=\rjet {\rm sin} \theta {\rm cos} \phi, \\
y=\rjet ({\rm cos} \theta {\rm sin} \thetaobs + {\rm cos} \thetaobs {\rm sin} \theta {\rm sin} \phi),\\
z=\rjet ({\rm sin} \theta {\rm cos} \thetaobs - {\rm sin} \thetaobs {\rm sin} \theta {\rm sin} \phi),
\end{array}
\end{equation}

where $\theta \in [0, \thetajet]$ and $\phi \in [0, 2\pi]$, with $\thetajet$ defined as the
jet half-opening angle.
The distance of a point $P$  on the top-hat surface  to the observer located at position $O$ with coordinates (0,0,$d$) is

\begin{equation}
    r = \sqrt{d^2 - 2\rjet\cos\thetaobs\cos\theta+2\rjet d \sin\thetaobs\sin\theta\sin\phi + \rjet^2}.
    \label{total_distance}
\end{equation}

The cosine of the angle between the radial velocity vector and the line from $P$ to the observer ($PO$) is 

\begin{equation}
    \cosalpha=\frac{-\rjet + d~ \costhobs \costheta - d~ \sinthobs \sintheta ~\sinphi}{\sqrt{d^2 + \rjet^2 -2d \rjet \costhobs \costheta + 2 d~ \rjet \sinthobs \sintheta ~\sinphi}},
    \label{cosalpha}
\end{equation}

\noindent
which becomes for $r << d$

\begin{equation}
    \cosalpha \approx \costhobs~\costheta-\sinthobs~\sintheta~\sinphi.
     \label{cosalpha_approx}
\end{equation}

The cosine of the angle formed by $PO$ and the $z$-axis is 
\begin{equation}
    \cos\omega = \frac{1}{r}\left(d - \rjet\cos\thetaobs\cos\theta + r \sin\thetaobs\sin\theta\sin\phi\right),
\end{equation}

\noindent
and is $\approx 1$ under the same condition.

The specific intensity in the observer frame is obtained by a Lorentz transformation along with the cosmological correction term and is given by

\begin{equation}
 \Ijet(\vv{k}_{\rm obs}, E_{\rm obs})=  \frac{\Ijet(\vv{k}_{\rm jet}, E_{\rm jet}) D_{\rm f}^3}{(1+z)^3},
 \label{transform_intensities}
\end{equation}    
 \noindent
 where
\begin{equation}
    E_{\rm jet}=\frac{(1+z) E_{\rm obs} }{D_{\rm f}},
    \label{ejet_vs_eobs}
\end{equation}
\noindent
while
\begin{equation}    
 D_{\rm f} = 1/\Gamma(1-\beta\cos\alpha),
 \label{doppler}
\end{equation}

\noindent
is the usual Doppler factor, $\Gamma$ is the bulk Lorentz factor and $\beta$ is the fluid's velocity 
in units of the speed of light. We consider isotropic emissivity in the jet comoving frame
so that  $I_{\rm jet}(\vv{k}_{\rm jet}, E_{\rm jet})=  I_{\rm jet}(E_{\rm jet})$. 


\noindent
Let us now consider an element area on the jet-surface $dA = \rjet^2 \du ~\dphi$, centered at coordinates [$u,\phi]$, where $u={\cos(\theta)}$.
The effective solid angle subtended by this element area
is 

\begin{equation}
 \domegaeff_{\rm i,j}=(1+z)^4 \frac{\rjet^2  ~\cosalpha}{\dl^2}\du~\dphi,
 \label{domega}
\end{equation}

where we used the relation $\dl=(1+z)^2\da$ between
the angular diameter distance and the luminosity
distance, while the term $\cosalpha$ accounts for the
effective projected area.

Combining equations (\ref{transform_intensities})
and (\ref{domega}) the flux received by the observer at $O$ from the surface area $dA$ is 

\begin{equation}
dF(\eobs) = \frac{(1+z)  ~\rjet^2 }{\dl^2} I_{\rm jet} (\ejet) \Df^3~\cosalpha~ \du~ \dphi.
    \label{df_form}
\end{equation}

The jet is assumed to emit a single pulse in the stationary (redshift-corrected) rest-frame so that
$L(t_{*})=\delta (t_{*}-t_{*,0})$: because of the curvature effect,  the pulse signal duration in the observer frame for $\thetaobs \leq \thetajet$ is 

\begin{equation}
     t^{\rm on}_{\rm p}=(1+z) \frac{\rjet}{c}[1-\cos(\thetaobs+\thetajet)],
\label{tspa_on}
\end{equation}
\noindent
while for $\thetaobs > \thetajet$ it is
\begin{equation}
 t^{\rm off}_{\rm p}=(1+z) \frac{2 \rjet}{c} {\rm sin}\thetaobs {\rm sin}\thetajet.
    \label{tspa_off}
\end{equation}

\noindent
In the remainder of the paper we will label as on-axis
and off-axis the events belonging to the first and second
case, respectively.\footnote{Note that some authors label as off-axis the events for which $\thetaobs > 0^{\circ}$, no matter on the value of $\thetajet$.} 

\noindent
The flux from the whole top-hat surface averaged over the pulse time $t_{\rm p}$  is obtained as 

\begin{equation} 
    F(\eobs)=  \frac{(1+z)  ~\rjet^2 }{\dl^2~ (\tp/{\rm s})} \int^{2\pi}_0 \int^{1}_{\rm u_c} I_{\rm jet} (\ejet) \Df^3 \cosalpha ~\du ~\dphi,
\label{angular_integral_flux}
\end{equation}
 
 \noindent
 with  $u_{\rm c}=\costhjet$.

\noindent
The isotropic equivalent luminosity is computed
with the relation

\begin{equation}
    \liso  = 4\pi D_{\rm L}^2 (1+z) \int_{E_1/(1+z)}^{E_2/(1+z)} F [(1+z) \eobs] d\eobs,
    \label{liso}
\end{equation}
\noindent
and the fluence as

\begin{equation}
    \eiso=\frac{\Delta t_{\rm obs}}{(1+z)}\liso.
    \label{eiso}
\end{equation}

\noindent
The code modularity allows to choose among different models for the emissivity in the comoving frame; we consider first a smoothly broken-powerlaw (SBPL) in the form

\begin{equation}
I(E)= K\left(\frac{E}{E_0}\right)^{-p_1} \left\{\frac{1}{2}\left[1+ \left(\frac{E}{E_0}\right)^{1/\delta} \right] \right\}^{(p_1-p_2)/\delta},
\label{sbpl_model}
\end{equation}

\noindent
with $K$ in units of erg~cm$^{-2}$~s$^{-1}$~keV$^{-1}$~ster$^{-1}$ and the parameter $\delta$ determines the smoothness of the transitions
between the two PL regimes with index $p_1$ and $p_2$, and the change of slope occurs in the energy interval $E_1 - E_2$ such that

\begin{equation}
    {\rm log}_{10} \frac{E_2}{E_0}={\rm log}_{10} \frac{E_0}{E_1} \sim \delta.
\end{equation}

The second model is a cut-off powerlaw (CPL) defined as

\begin{equation}
    I(E)=K (E/{\rm keV})^{-q} e^{-E/E_{\rm cut}}.
    \label{cpl_model}
\end{equation}

Finally, for a thermal component we use a simple blackbody (BB) function

\begin{equation}
I(E)= K \frac{(E/{\rm keV})^3}{e^{E/kT_{\rm bb}}-1}.
\label{bb_model}
\end{equation}

 The normalization $K$ is allowed to be free in the case of BPL and CPL models, while in the BB case it is dictated by its thermodynamical limit value  $5\times 10^{22}$ erg~cm$^{-2}$~s$^{-1}$~keV$^{-1}$~ster$^{-1}$. 
In this paper, all presented results have been obtained assuming
a comoving-frame SBPL spectrum.

\section{Results}
\label{result}

\subsection{Constant $\Gamma$-factor}

\begin{figure*}
 \hspace{-0.2in}\includegraphics[scale=0.5]{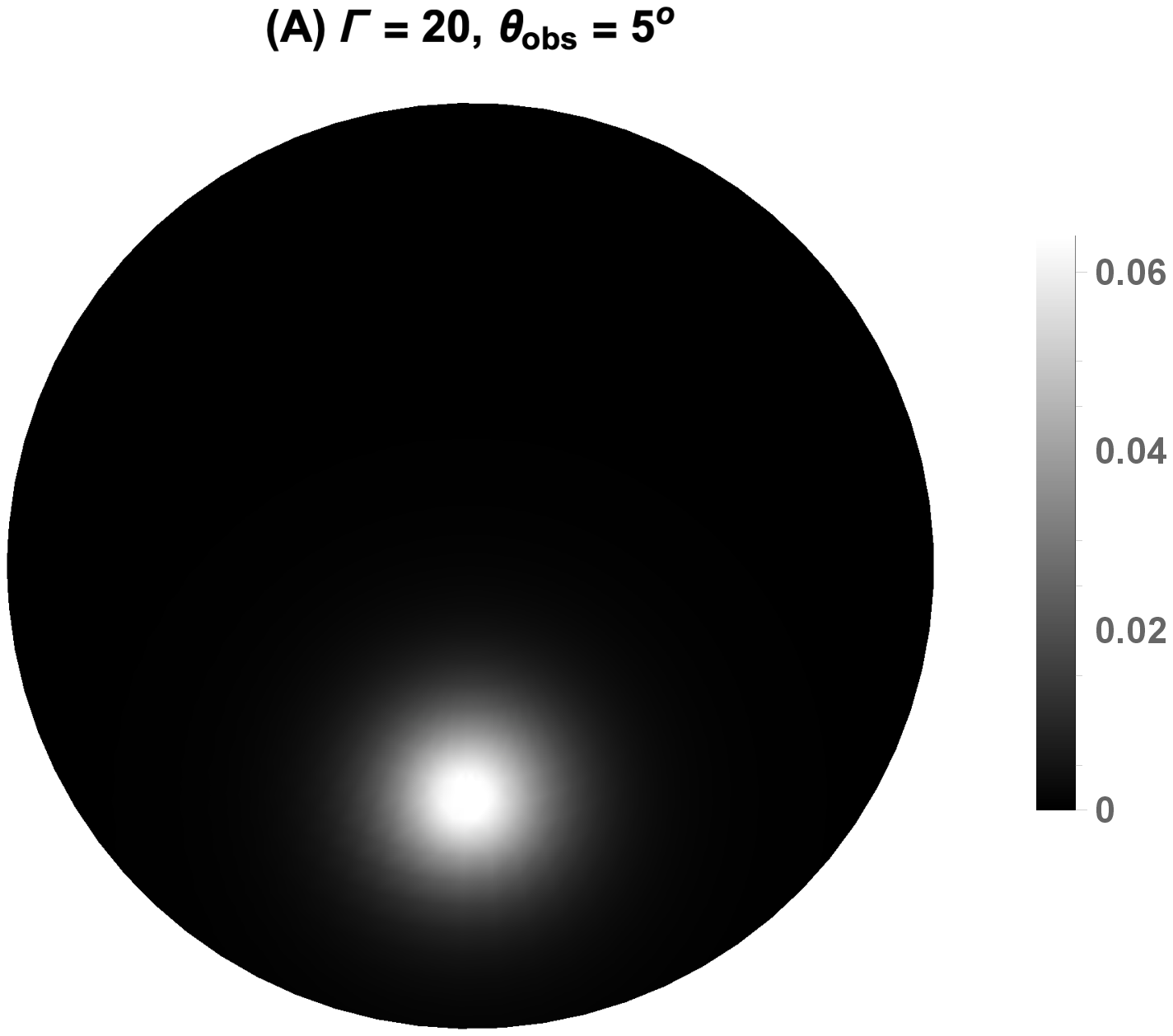} 
  \hspace{0.1in}
 \includegraphics[scale=0.5]{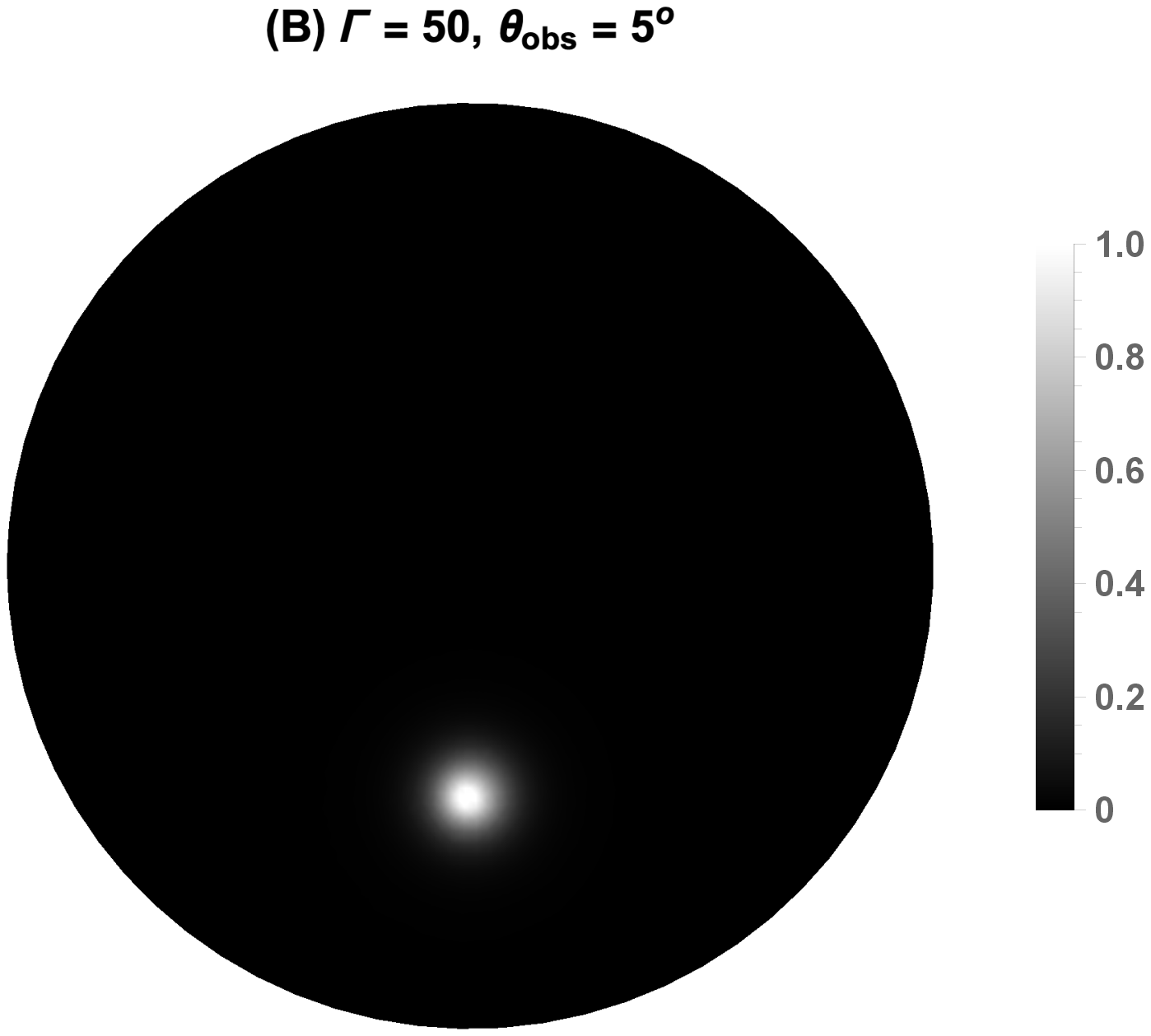}
 \includegraphics[scale=0.5]{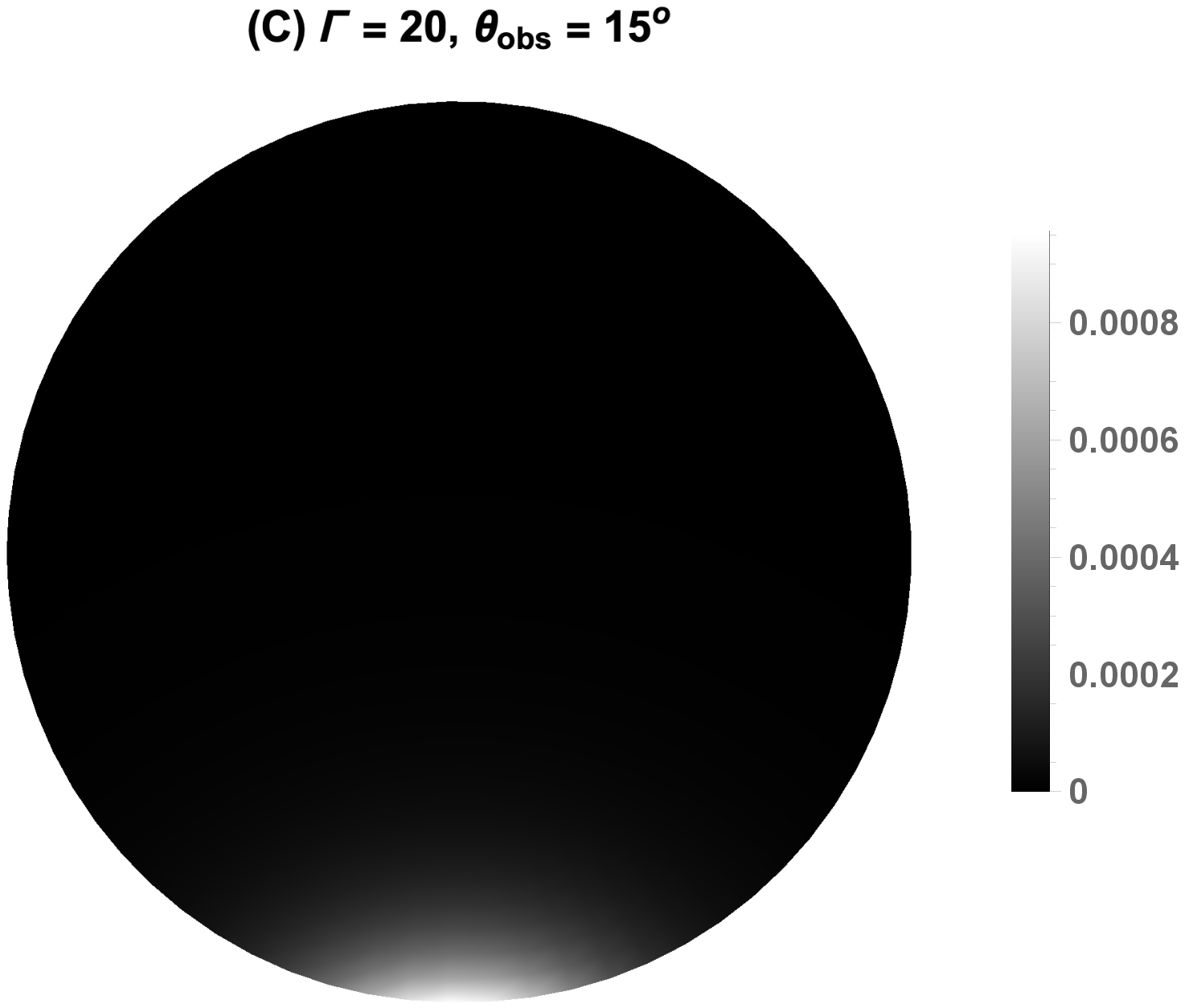}
 \includegraphics[scale=0.5]{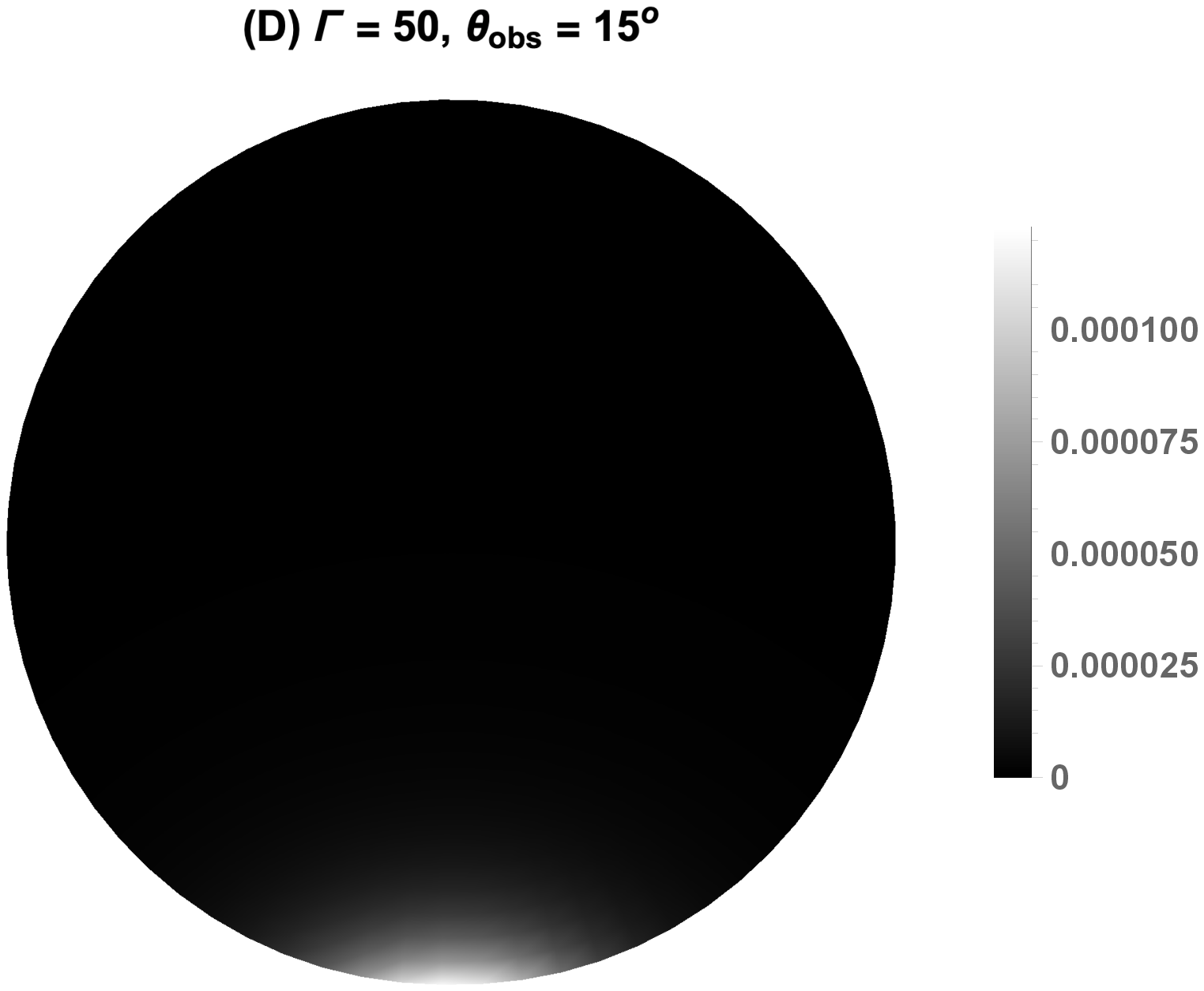}
\caption{Top-surface brightness along the $xy$-plane of a jet with opening angle $\thetajet=10^{\circ}$ and different values of the $\Gamma$ 
factor and observer's viewing angle. \emph{Top-left}: (A) $\Gamma=20, \thetaobs=5^{\circ}$,~\emph{Top-right}: (B) $\Gamma=50, \thetaobs=5^{\circ}$,~\emph{Bottom-left}: (C) $\Gamma=20, \thetaobs=15^{\circ}$,~\emph{Bottom-right}: (D) $\Gamma=50, \thetaobs=15^{\circ}$.}
\label{surface_brightness}
\end{figure*}

In order to emphasize the importance of the observer viewing angle, we show in Figure~\ref{surface_brightness} the observed surface brightness of the jet for four different combination of $\Gamma = 20, 50$ and $\thetaobs = 5^{\circ}, 15^{\circ}$, while we fix $\thetajet = 10^{\circ}$. The values shown in the colour bars are not in absolute units, but should be used to compare the relative brightness between different cases. 
The relative brightness drops by several orders of magnitude from a viewing angle inside the jet cone ($\thetaobs < \thetajet$, upper panels) to outside the jet cone ($\thetaobs > \thetajet$, lower panels).  For the former case, i.e., $\thetajet<\thetaobs$, the lateral spreading becomes higher and the relative brightness becomes lower as $\Gamma$ decreases (compare panel A with panel B). However, for $\thetajet>\thetaobs$ (see the lower panels), we note an opposite trend. Here, the brightness is relatively higher by a factor of $\sim8$ for lower $\Gamma$ (compare panel C with panel D). 
It is evident that we cannot always associate brighter GRBs with higher $\Gamma$. The observer needs to be within the jet cone for that to happen, otherwise a reverse variation will be seen.

\begin{figure*}
  \includegraphics[scale=0.54]{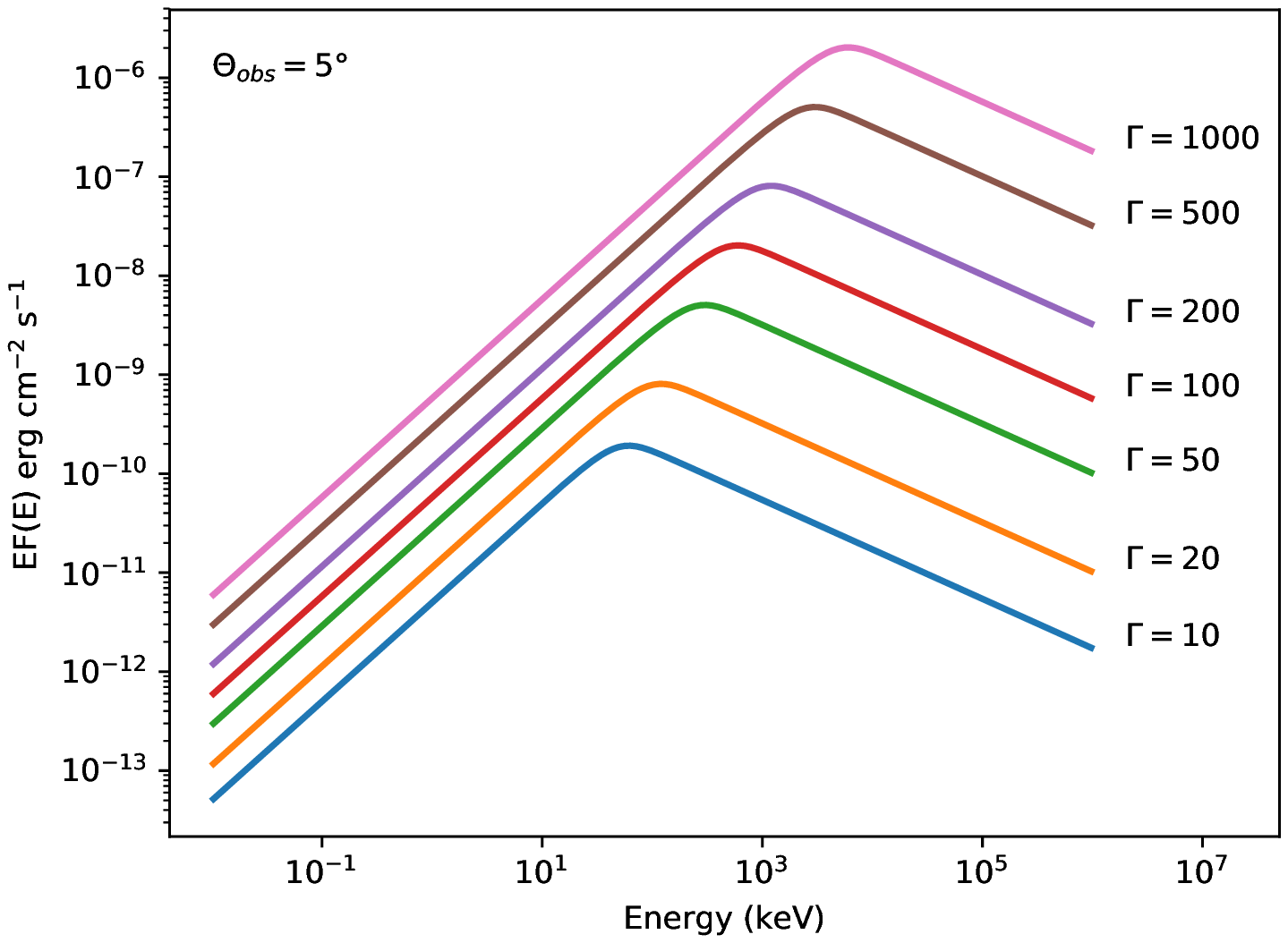}
  \includegraphics[scale=0.54]{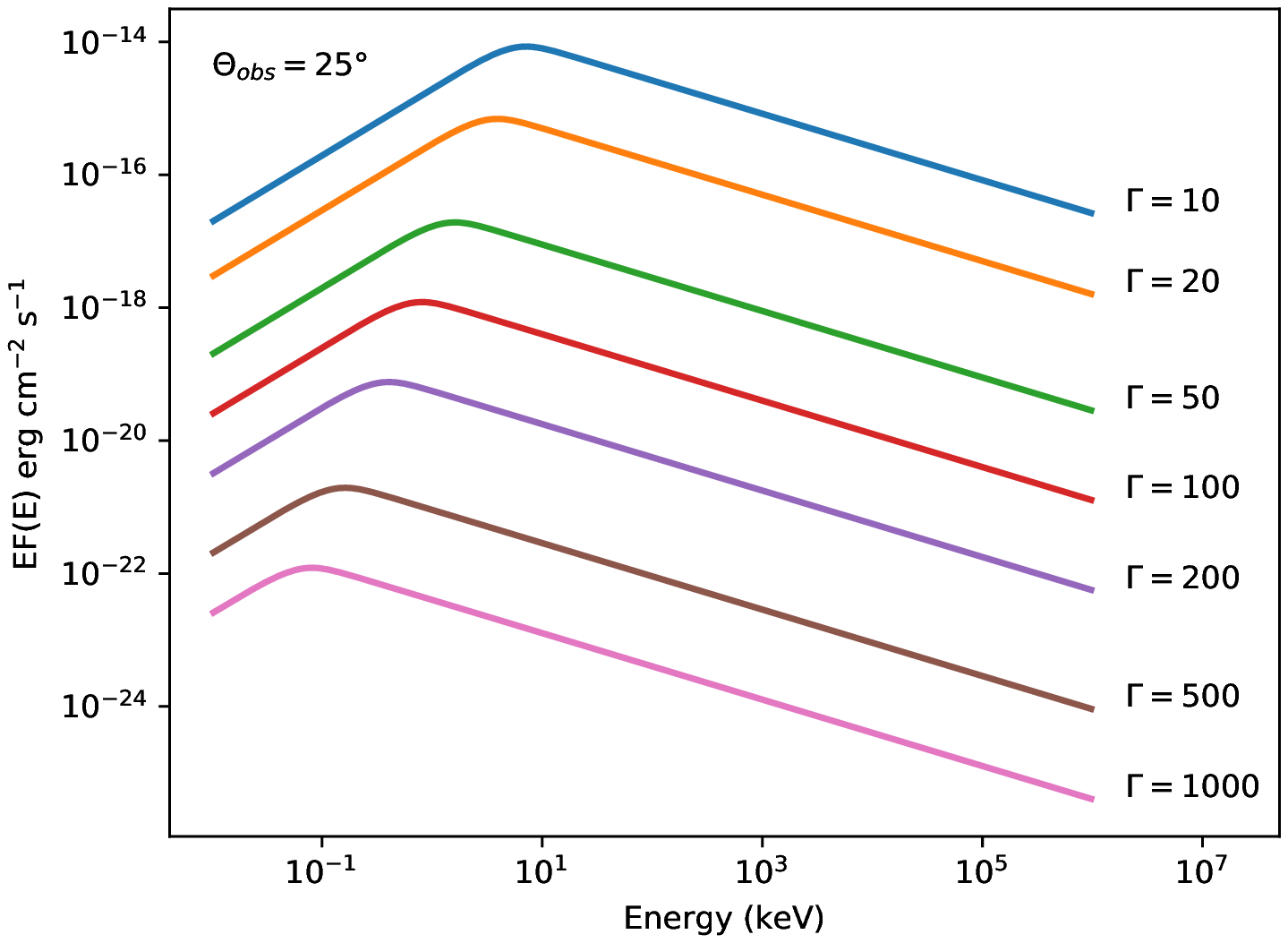}
\caption{Simulated spectra  with $\Gamma$ varied in the range $10-1000$ for a jet with $\thetajet=10^{\circ}$, $\rjet=10^{12}$ cm and a SBPL emissivity law in the comoving frame with $E_0=10$ keV, $p_1=0$, $p_2=1.5$ and $\delta=0.2$.
The two cases for  $\thetaobs=5^{\circ}$ (on-axis) and   $\thetaobs= 25^{\circ}$ (off-axis) are reported.}
\label{spectra_vargamma}
\end{figure*}

We then consider the observed spectra and first focus on the cases for different values of the $\Gamma$ factor for both the on-axis and off-axis case. The results are shown in Figure \ref{spectra_vargamma}.
The simulations have been performed assuming a jet with $\thetajet=10^{\circ}$ and radius $\rjet=10^{12}$ cm, and a SBPL emissivity law 
in the comoving frame with $p1 = 0$, $p2 = 1.5$, $E_{\rm 0} = 10$ keV and $\delta=0.2$.

\begin{figure*}
\includegraphics[scale=0.54]{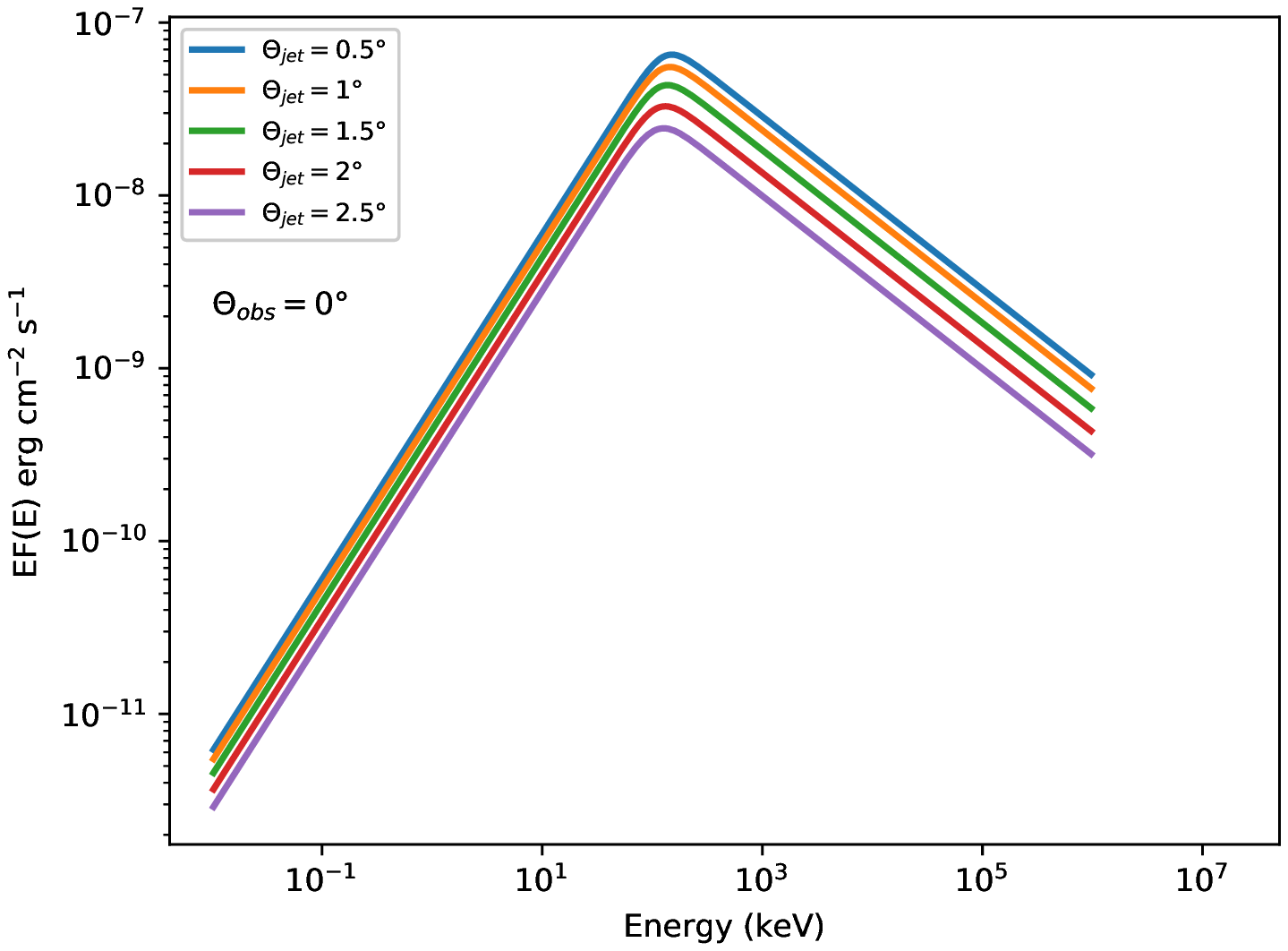} \includegraphics[scale=0.54]{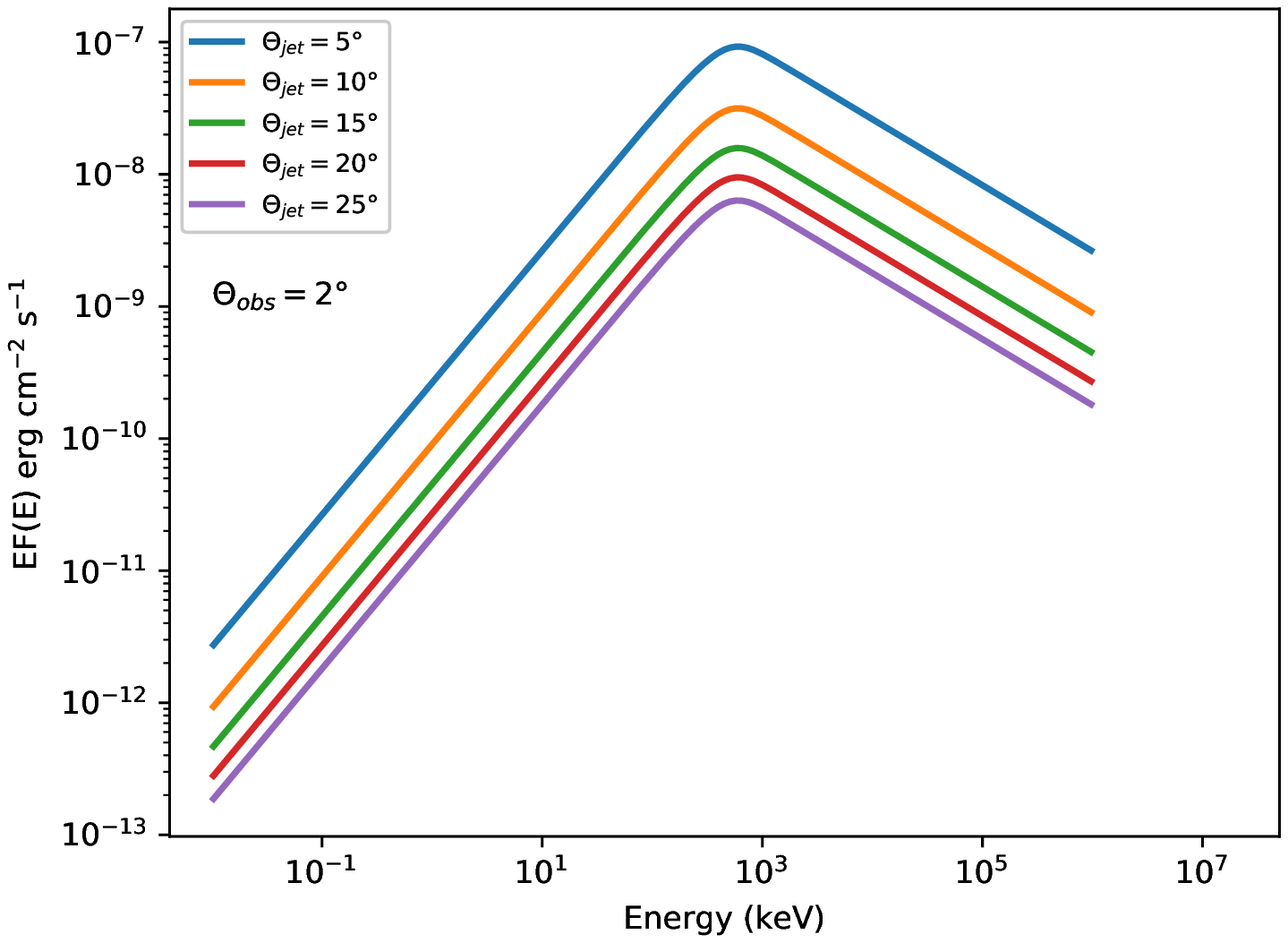}
\includegraphics[scale=0.54]{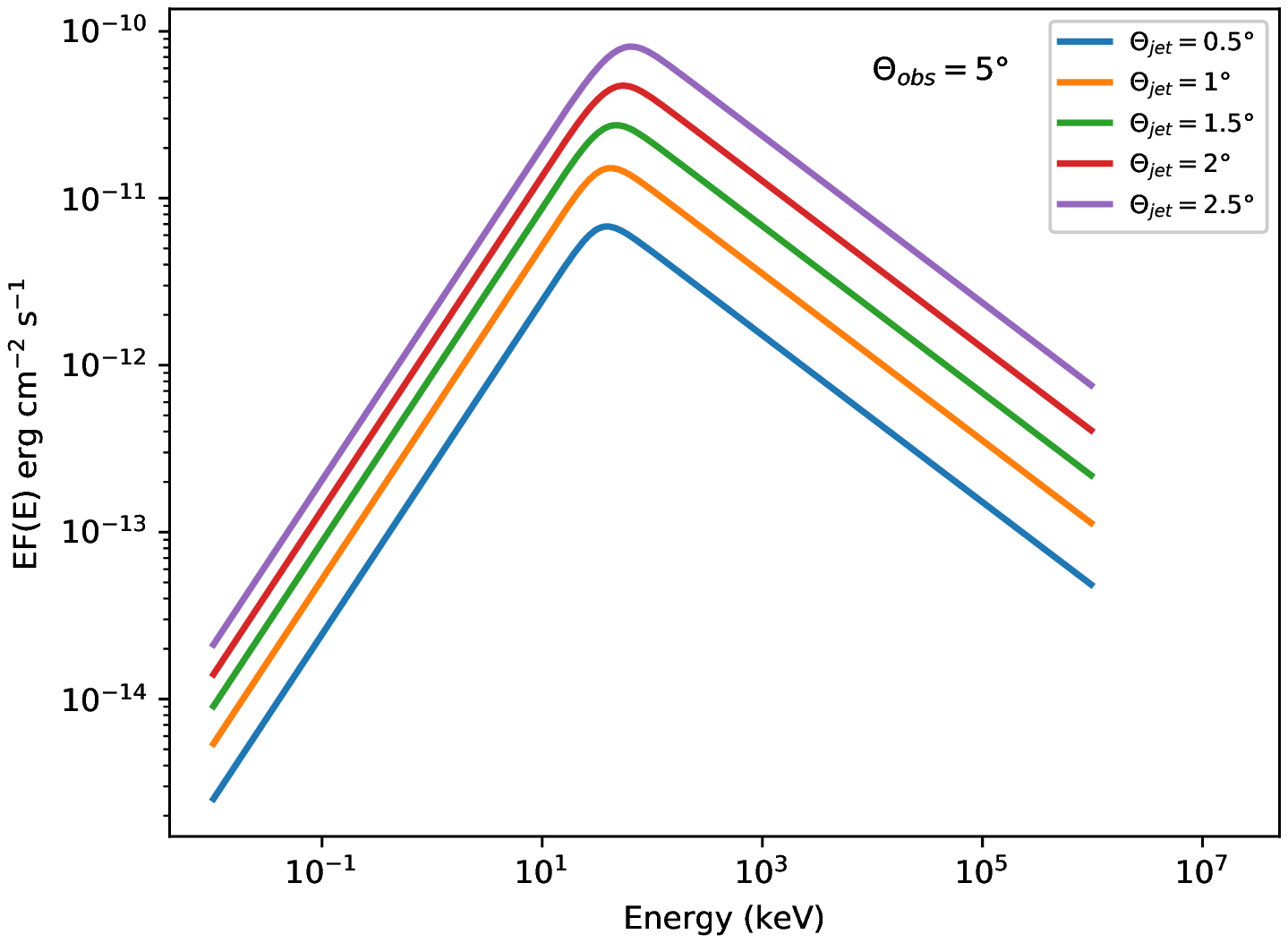} \includegraphics[scale=0.54]{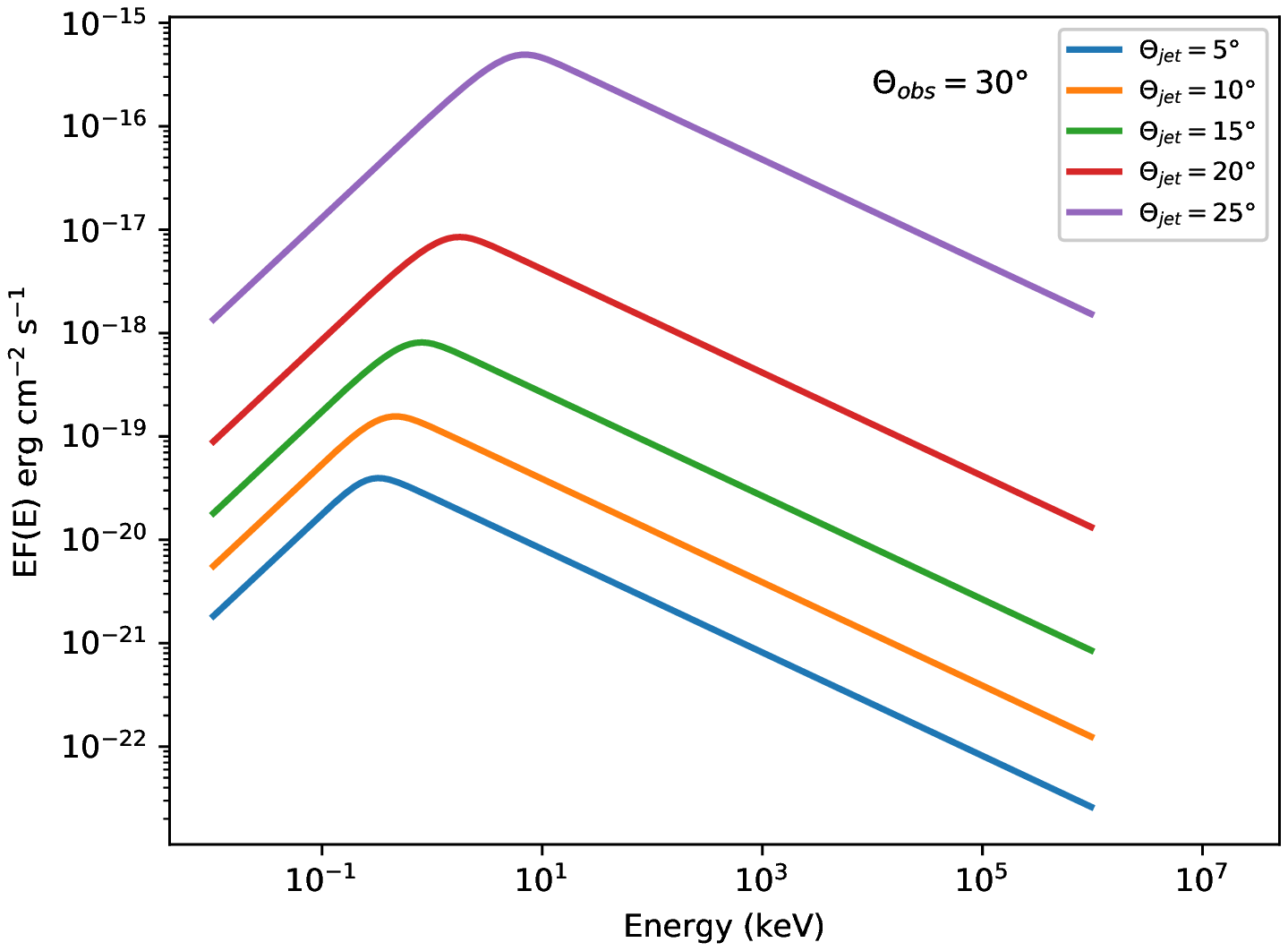}
\caption{Simulated spectra for different values of the jet opening angle $\thetajet$. \emph{Top panels}: on-axis case with $\Gamma=20$ such that $\thetajet < 1/\Gamma$ (\emph{left}) and  $\Gamma=100$ with $\thetajet >> 1/\Gamma$  (\emph{right}). 
\emph{Bottom panels}: off-axis case. 
The jet radius is $10^{12}$ cm, while the parameters of the comoving-frame SBPL emissivity law are the same of Figure \ref{spectra_vargamma}.}
\label{spectra_varthobs}
\end{figure*}


For the cases $\thetaobs<\thetajet$, the spectral peak and flux are positively correlated with the $\Gamma$. On the contrary, for  observer's viewing angle outside the cone ($\thetaobs>\thetajet$), 
a reverse variation is noticed. 
It is also worth pointing the significant drop
in the received flux for off-axis events, confirming
that most of GRBs belonging to this class are likely  below the threshold sensitivity of current instruments, unless either the $\Gamma$ factor is low or they
are neighbour (see Section \ref{amati_relation}).

The second parameter we investigated is the jet opening angle $\thetajet$, and results are presented in Figure \ref{spectra_varthobs}: for $\thetaobs < \thetajet$ and $\Gamma$ such that
$\thetajet >> 1/\Gamma$, all the spectra are equivalent, with their normalization rescaled by the pulse duration $\tp$ 
(see equations [\ref{tspa_on}] and [\ref{angular_integral_flux}]).
Note that this renormalization unavoidably comes from the
model definition, but actually the bulk of the flux comes from a region of angle $\theta \sim 1/\Gamma << \thetajet$ centered towards the observer's viewing angle, which is  independent
on $\thetajet$.  This can be seen noticing that the observed
peak energy $\ep$ remains unchanged, and confirms the well-know result that for $\thetajet > 1/\Gamma$ a radially symmetric jet is indistinguishable from
an expanding sphere \citep[e.g.,][]{rhoads97, salafia16}.

Typical GRB Lorentz factors are $\Gamma \ga 100$
\citep{meszaros06}, and the effect of
observer's viewing angle in the on-axis case would require very collimated jet opening angles $\thetajet \la 1^{\circ}$. This leads in turn to  two consequences at the observational level; firstly, such very narrow jets (if present) have a
very low probability of detection, second,
the dynamical range of observer's viewing angle
with $\thetaobs \leq \thetajet$ would be so
narrow to essentially suppress any detectable
effect.\\
Nevertheless, for illustrative purposes we report in Figure \ref{spectra_varthobs} (top-left panel) the case of a moderate Lorentz factor $\Gamma=20$
and values of $\thetajet \la 1/\Gamma$ for a fully
on-axis observer ($\thetaobs=0^{\circ}$).
Here, there is not just a simple spectral rescaling
due to $\tp$ but also a progressive decreases
of $\ep$ as $\thetajet$ increases.\newline
The situation changes when looking outside the jet cone
no matter whether $\thetajet$ is lower or higher than $1/\Gamma$: in Figure \ref{spectra_varthobs} (bottom panels) we report simulated spectra again for $\thetajet < 1/\Gamma$
and $\thetajet >> 1/\Gamma$.
In this case the trend is always the same, with $\ep$ increasing
as $\thetajet$ does as well. This is understood because
for off-axis events, the dominant contribution to the observed flux always 
comes from the region around the bottom border of the jet (see also Fig. \ref{surface_brightness}).
For fixed observer's viewing angle, as $\thetajet$ increases
the Doppler factor from this area increases, leading in turn o higher observed values of $\epobs$.
We also claim that using only spectral fitting, the jet opening angle cannot be measured in the GRB prompt phase for on-axis events apart for the unlikely
case $\thetajet \la 1^{\circ}$ (see top-left panel of Figure~\ref{spectra_vargamma}), while in principle
it could be done for off-axis events (if detectable) where
the effect of $\thetajet$ in the observed spectra is much more enhanced.
Note that we do not present results for varying
values of $\rjet$ as the pulse-average flux
simply rescales linearly with it  
(equation [\ref{angular_integral_flux}]).
Concerning the parameters of the comoving-frame SBPL
(equation [\ref{sbpl_model}]),
apart for the low-energy and high-energy slopes
(depending on $p_1$ and $p_2$ respectively) and
the smoothness of the transition at the peak energy $E_0$
(depending on $\delta$), we point out that the both the flux
and $\ep$ linearly scales with $E_0$, for fixed
values of $\thetaobs$ and jet parameters.

\subsection{Structured jet with variable $\Gamma$-factor}

In a more complicated scenario, the GRB jet may have an angular dependence $\Gamma(\theta)$ of the bulk Lorentz factor
(see references in Section \ref{intro}). 
Numerical hydrodynamic simulations have been performed e.g., by \cite{zhang03}, who showed angular variation of density, flux and $\Gamma$. Later,  \cite{lundman13} (hereafter L13) used an empirical form to characterize the derived $\Gamma$-profile.  In this section, we explore the effects of a non-constant $\Gamma$ on the observed properties of our model. In particular, we adopt the analytical function proposed by L13, with slight modification, where we explicitly write the dependence of $\Gamma$ on $\theta$:


\begin{equation}
\Gamma=\gmin + \frac{\gmax-\gmin}{\sqrt{(\theta/\thetajet)^{2p}+1}},
\label{eq_gamma_vs_theta}
\end{equation}

\begin{figure}\centering
\includegraphics[width=\columnwidth]{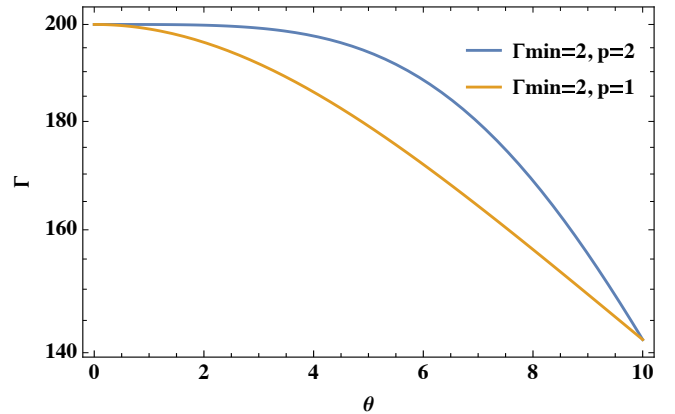} 
\caption{Dependence of the $\Gamma$-factor as a function of $\theta$ (see equation \ref{eq_gamma_vs_theta}) with $\thetajet=10^{\circ}$, $\gmax=200$,  $\gmin=2$ and two different values of $p$.}
\label{fig_gamma_vs_theta}
\end{figure}

where $\gmax$ is the maximum value of the Lorentz factor at the center of the jet. It is important to keep in mind that $\gmin$ \emph{is not} the value of $\Gamma$ for $\theta=\thetajet$, but it represents an asymptotic value of $\Gamma$ at the outer layer, for instance in the presence of a sub-relativistic surrounding cocoon.
To better show this effect, in Figure \ref{fig_gamma_vs_theta}, we report the behaviour $\Gamma$ as a function of $\theta$ for the case $\gmax=100$, $\gmin=2$ and two different values of the index $p$.

\begin{figure}\centering
\includegraphics[width=9cm, height=8cm]{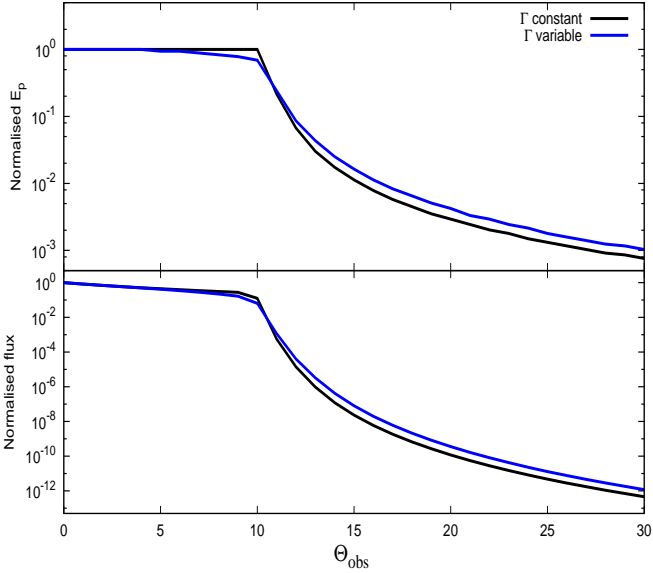} 
\caption{Dependence of the observed $\ep$ and flux as a function of the  observer  viewing angle $\thetaobs$ for a constant Lorentz factor $\Gamma=100$ and for $\Gamma$ obeying the law in equation (\ref{eq_gamma_vs_theta}) with $\gmax=100$, $\gmin=1.1$ and  $p=1$. We assumed $\thetajet=10^{\circ}$, and the values are normalized to the case $\thetaobs=0^{\circ}$.}
\label{epeiso_vs_thobs}
\end{figure}

To investigate the observational effects of the $\Gamma$-stratification, we show in Fig. \ref{epeiso_vs_thobs} the observed $\ep$ and flux for different values of the observer's viewing angle, with $\thetajet=10^{\circ}$ and for the case of constant  and variable $\Gamma$-factor, respectively. The quantities are normalised to the value at $\thetaobs=0^{\circ}$ as we are interested to consider relative rather than absolute variations. For the first case, we choose $\Gamma=100$, while for the second case the chosen parameters are $\gmax=100$, $\gmin=1.1$ and $p=1$ (see equation~\ref{eq_gamma_vs_theta}). 

Let us first discuss the case of $\Gamma$-constant: for $\thetaobs \leq \thetajet$, $\ep$ and remains constant,
while the pulse-average flux simply rescales with $\tp$,
as defined in Equations (\ref{tspa_on}) and (\ref{tspa_off}). This is easily understood in view of the considerations made in previous
section when $\thetajet > 1/\Gamma$ -- the strong beaming effect suppresses
any dependence of the spectral shape on $\thetaobs$.
On the other hand, when $\thetaobs > \thetajet$, the strong boosting of the photons along the direction of motion, without any light-of-sight intercepting the jet top-hat,
causes a strong drop in the observed flux, which progressively decreases as $\thetaobs$ increases. The main contribution comes here from the bottom part of the jet surface, as previously outlined (see Figure \ref{surface_brightness}). 

The result changes slightly when considering a variable $\Gamma$-factor: when $\thetaobs \leq \thetajet$, both $\ep$ and the flux are mostly dictated by
the value $\Gamma(\thetaobs)$ which is lower than $\Gamma(0)$.
At the observational level however, a structured jet viewed at an angle $\thetaobs$
is indistinguishable from a $\Gamma$-constant jet with $\Gamma=\Gamma(\thetaobs)$.
 On the contrary, when $\thetaobs > \thetajet$ both $\ep$ and
 the flux are higher than the case of constant $\Gamma$, because
 at the jet border from which most of the emission comes
 the $\Gamma$-factor achieves its minimum, and the Doppler
 boosting out of the direction towards the observer is less pronounced.

The situation may be of course different in the more complicated scenario of a structured jet as reported e.g. in \cite{zhang03}, with a surrounding  sub-relativistic cocoon,  but this effect which results from magneto-hydrodynamical situations cannot be taken into account by our model.

In the version released for the \xspec\ package we did not
implement the dependence of $\Gamma$ on the polar angle, to avoid
having a too-high number of degrees of freedoms 
(additional $\Gamma_{\rm min}$ and $p$, see equation \ref{eq_gamma_vs_theta}), which actually overcomes the number
of observational spectral parameters.

\section{The $\ep-\eiso$ relation for on-axis and off-axis events}
\label{amati_relation}

\begin{figure}
 \includegraphics[scale=0.7]{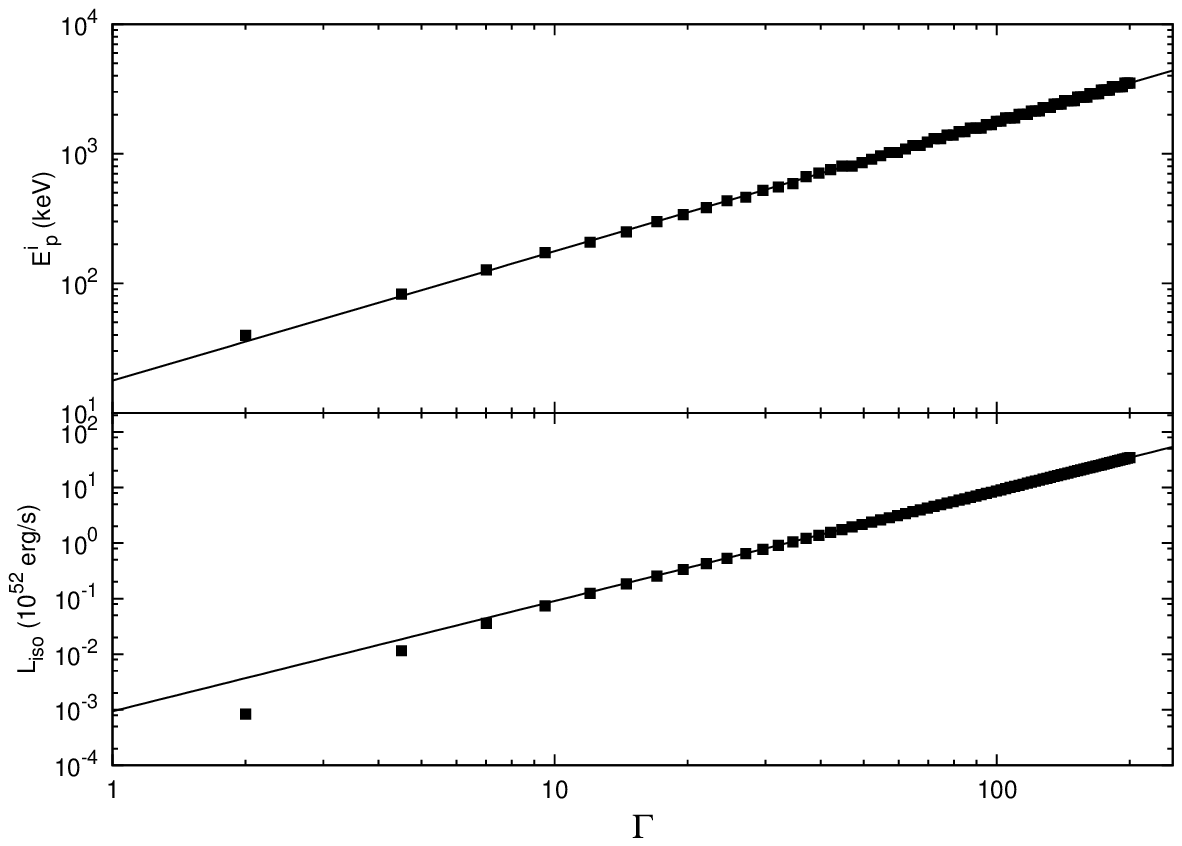} \hspace{0.1in}
 \includegraphics[scale=0.7]{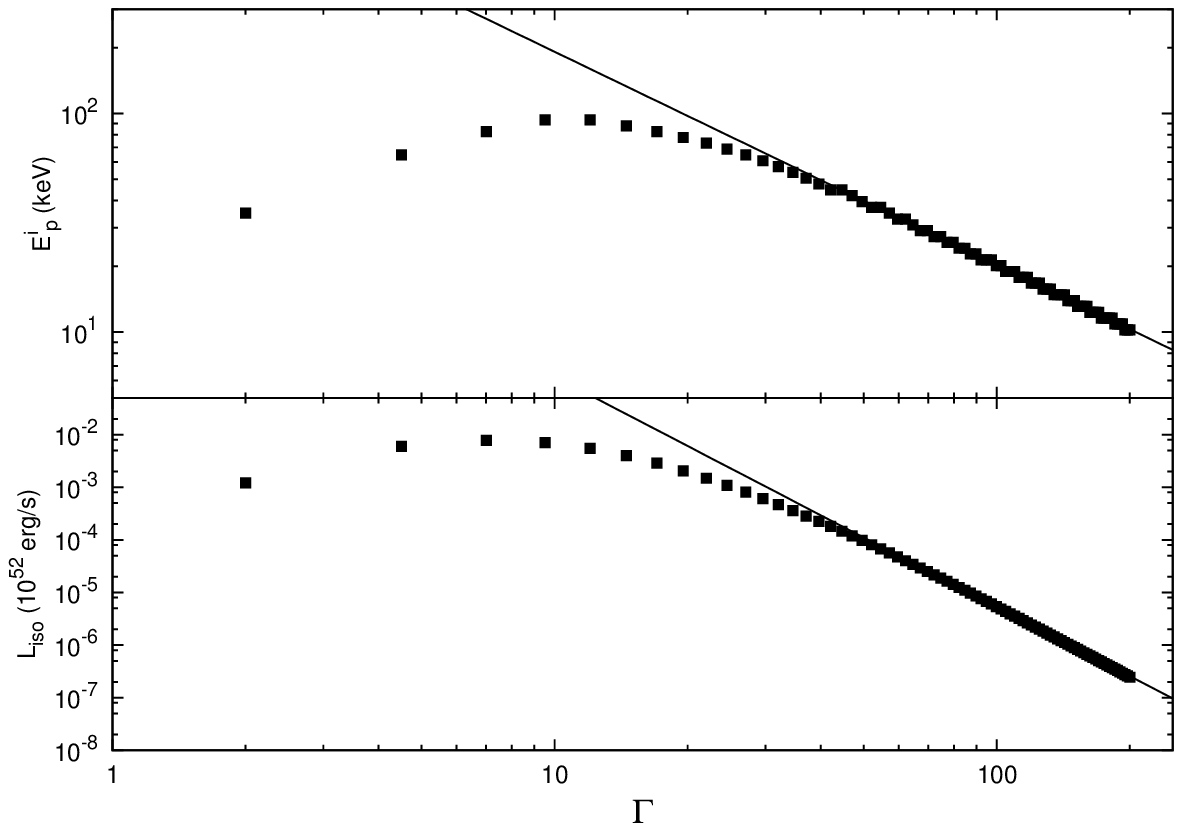}
\caption{Behaviour of $\epi$ and $\liso$ as a function of the Lorentz factor $\Gamma$, for the two cases of jet viewed on-axis (\emph{top panels}) and off-axis (\emph{bottom panels}). Jet parameters are  $\thetajet=10^{\circ}$ and $\rjet=10^{12}$ cm, while observer viewing angles are
$\thetaobs=5^{\circ}$ and $\thetaobs=15^{\circ}$. Overplotted to data are the best-fit powerlaw functions
for $\Gamma \ga 20$. For the on-axis
case the slopes are $\epi \propto \Gamma$ and $\liso \propto \Gamma^2$, for the off-axis case $\epi \propto \Gamma^{-1}$ and $\liso{} \propto \Gamma^{-4}.$}
\label{params_vs_gamma}
\end{figure}

The main parameter driving the observed spectral peak energies and fluxes in relativistic outflows is the $\Gamma$-factor of the emitting material \citep{dermer2004}.
We investigated this effect with a series of simulations at different  $\Gamma$-values (not depending on $\theta$), assuming the same SBPL spectrum in the comoving-frame with $E_0=10$ keV, $p_1=0$, $p_2=1.5$ and $\delta=0.2$ (equation \ref{sbpl_model}) and
for a jet radius  $\rjet= 10^{12}$ cm. The results are presented
in Figure \ref{params_vs_gamma}.

We obtain a clear dichotomy between the on-axis and off-axis cases; when $\thetaobs \leq \thetajet$ both $\epi$ and
 $\liso$ increases as $\Gamma$ does. In particular, for $\Gamma \ga 20$ we obtain $\epi \propto \Gamma$ and 
$\liso \propto \Gamma^2$, which in turn leads to the relation $\epi \propto \liso^{1/2}$.
On the contrary, when $\thetaobs > \thetajet$ the two observable parameters progressively increase for moderate values
of $\Gamma$ up to $\sim 40$, above which a decreasing powerlaw-like behavior occurs with 
$\epi \propto \Gamma^{-1}$ and $\liso \propto \Gamma^{-4}$-- consequently, one obtains $\epi \propto \liso^{1/4}$.

As a next step, we moved to the $\epi-\eiso$ plane in order to reproduce the AR with a possibly \emph{qualitative}
representation of its intrinsic dispersion, which is known to have a variance higher than that coming
from statistical uncertainties \citep{amati02, amati06}.
The data dispersion of the AR indicates that one or more physical and/or geometrical parameters play a role in addition to $\Gamma$ which
is claimed to have the leading role in dictating the observed slope $\epi \propto \eiso^{0.5}$.

Further considerations to point out for on-axis events
are the following:
\begin{itemize}
    \item[-]  $\epi$ values (Y-axis) are $\propto E_0~\Gamma$
    \item[-]  $\eiso$ values (X-axis) are $\propto E_0 ~K ~\Gamma^2~\rjet^2$ 
\end{itemize}
\noindent
Moreover, all the parameters must be combined in order to reproduce not only the observed $\epi-\eiso$ slope, but also the normalization which is of order of $\sim 100$ \citep{amati09}. 

For off-axis events, $\epi$ and $\eiso$  have the same dependence on $E_0$, $K$ and
$\rjet$, but with an inverse proportionality on
$\Gamma$ as shown above.

Based on the above considerations and constraints, 
we proceeded in the following way: let us define $G(P)=N(P_{\rm c}, \sigmap)$ as the normal gaussian distribution of a given parameter
$P$ with $P_{\rm c}$ and  $\sigmap$ its mean and standard deviation, respectively.

Performing simulations, we assumed $G(\Gamma)=N(100,50)$ as derived from observational estimations \citep[e.g.,][]{ghirlanda2012}. 
Sampling $\Gamma$-values from such a distribution poses 
constraints about the allowed range of $E_0$-values when comparing numerical results to observed values of $\epi$
which cluster, a-part from a couple of cases below 10 keV, in a range of values from few tens keV to few MeV \citep{amati09}.
With the assumed distribution of $\Gamma$, we found that 
a good  choice for the comoving-frame break energy is 
$G(E_0)=N(5,2)$. 

For the low-energy and high-energy index of the comoving-frame SBPL, we draw random values from  two distributions with  $G(p_1)=N(0,0.5)$ and $G(p_2)=N(1.5,0.5)$, respectively \cite[see][for a sample of best-fit parameters using data from \emph{BATSE} and \emph{Fermi/GBM} sample]{nava2011}.

The value of the jet half-opening angle $\thetajet$ is instead drawn from a uniform distribution in the range $5^{\circ}-20^{\circ}$, while the observer's viewing angle is sampled from a uniform distribution over cos$(\thetaobs)$ with
$0 \leq \thetaobs \leq \thetajet$ for on-axis
events, $\thetajet < \thetaobs < \thetajet + 20^{\circ}$
in the other case.
Finally,  for the product of jet radius and SBPL
normalization in such a way that $G(R^2_{12}/K_{20})=N(150,10)$, where $R_{12}$
and  $K_{20}$ are in units of $10^{12}$ cm and 
$10^{20}$ ergs~cm$^{-2}$~s$^{-1}$~keV$^{-2}$~ster$^{-1}$, respectively.

It is worth pointing that the independent sampling
of all parameters  implies a diagonal covariance matrix which  leads
in turn to a  variance of the data dispersion higher than that expected
if at least some physical quantities are correlated.
However, we are here interested in testing the $\epi-\eiso$ main trend
rather than its intrinsic dispersion, and the 
random parameter sampling has been adopted just to
simulate a \emph{qualitative} representation of the data dispersion.

\begin{figure*}
 \includegraphics[scale=0.5]{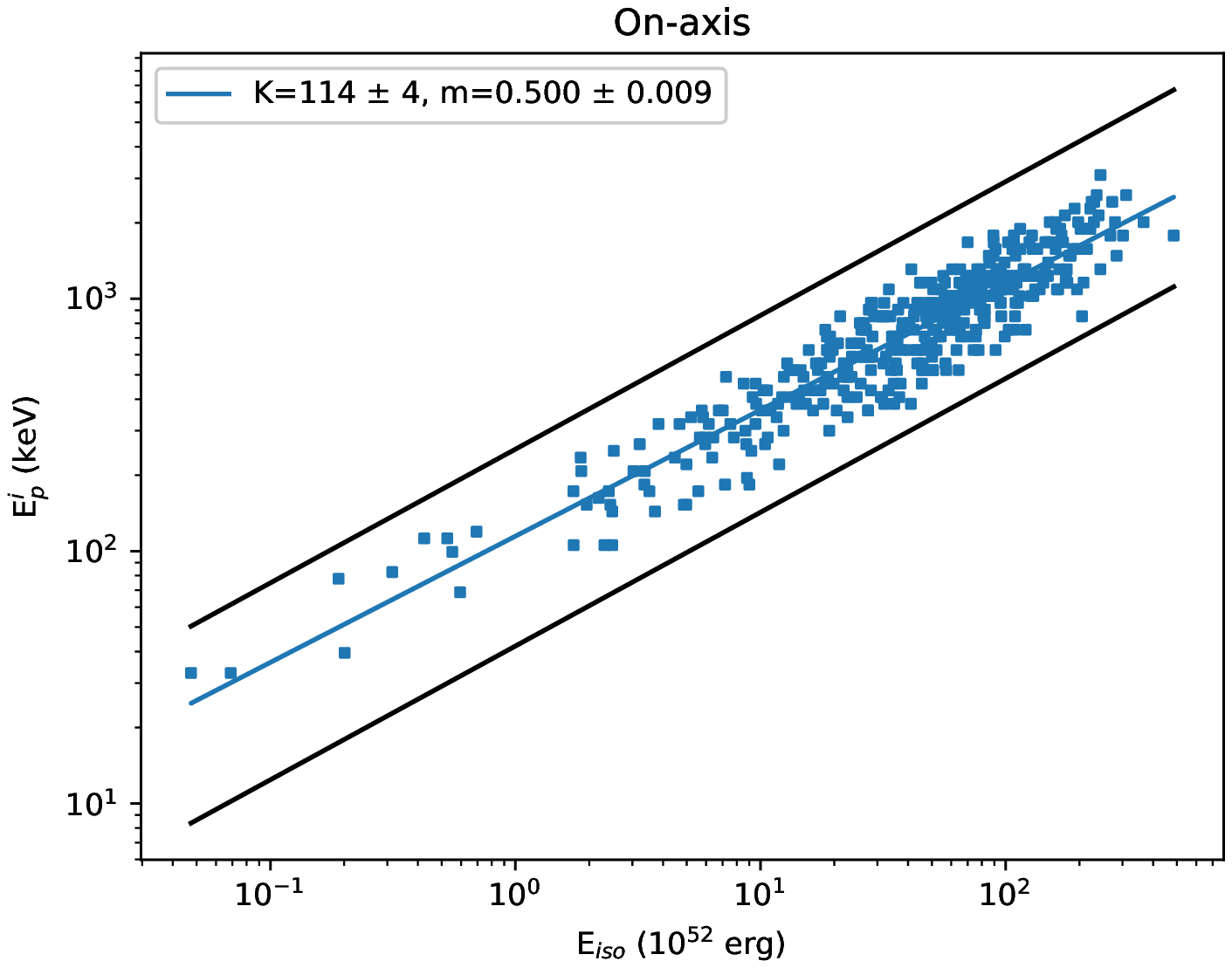} \hspace{0.1in}
 \includegraphics[scale=0.5]{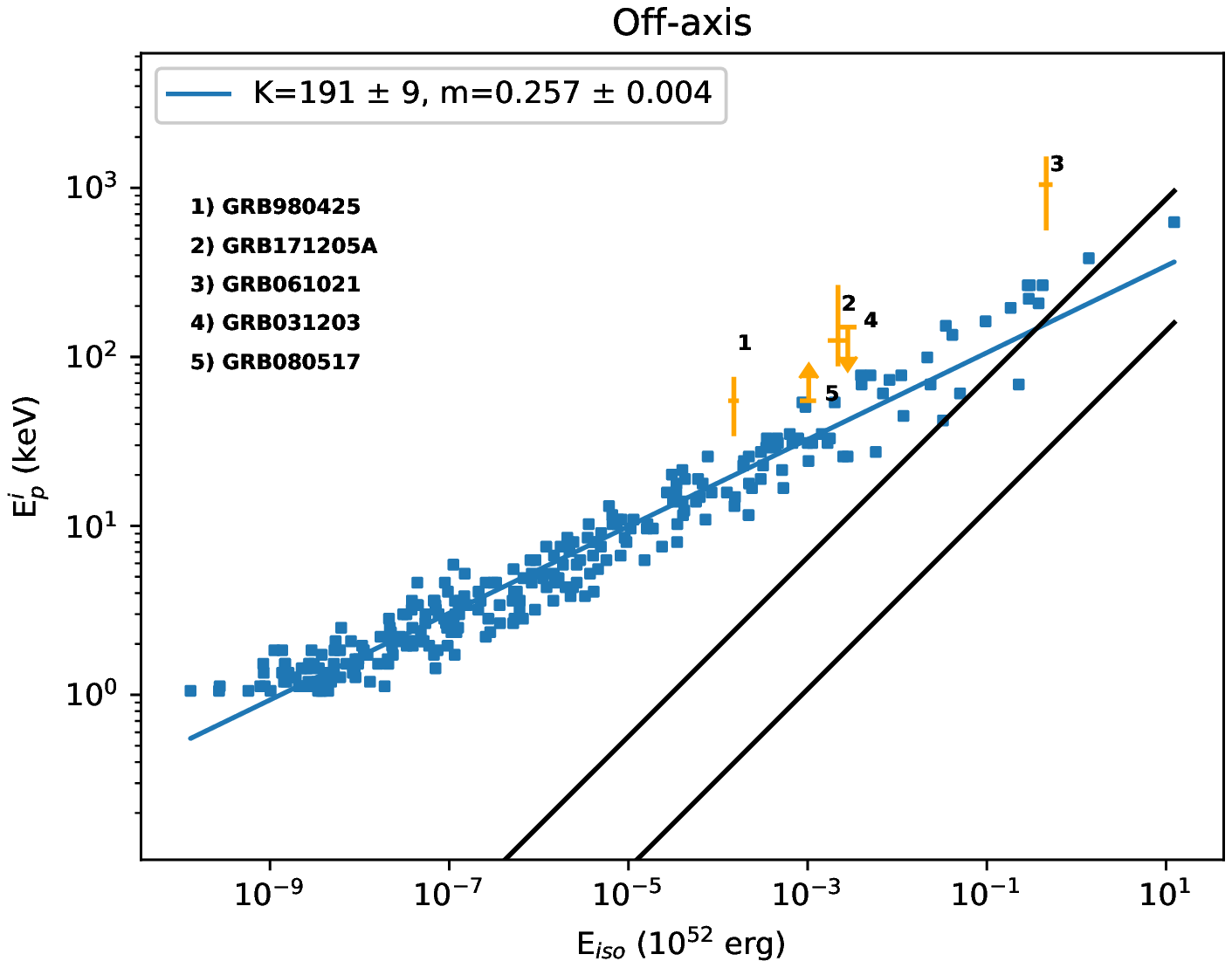}
\caption{Simulated $\epi-\eiso$ relation for events observed on-axis (\emph{left panel}) and off-axis (\emph{right panel}). References for true GRBs data are reported in Section \ref{amati_relation}. For GRB031203 and GRB080517 the upper and lower limits on $\epi$ respectively, are reported. The two continuous black lines correspond to the $\pm 2\sigma$ dispersion region of the observed $\epi-\eiso$ relation reported by \citet{amati08}.}
\label{epeiso_twocases}
\end{figure*}

The results are reported in Figure \ref{epeiso_twocases}:
as expected from the behavior of $\epi$ and $\liso$ as a function
of $\Gamma$ (see Fig. \ref{params_vs_gamma}), we obtain two different slopes
for the $\epi-\eiso$ relation for the two cases.
For on-axis events the index is $\sim 0.5$, and the dominant contribution to the 
data variance around the best-fit straight here comes 
from the random sampling of the comoving-frame SBPL parameters as well as
the product $R^2_{12}/K_{20}$. The jet half-opening angle and the
observer viewing angle play instead no role (see Figure \ref{epeiso_vs_thobs})
under the condition $\thetajet > 1/\Gamma$ whichis  always satisfied here.
Similar results have been obtained also by considering a structured jet according to equation (\ref{eq_gamma_vs_theta}).
In this case indeed, the net effect is to put an event 
observed at given angle $\thetaobs$ in the same  location of the $\epi-\eiso$ plane of events with constant $\Gamma$ viewed at any angle $\thetaobs < \thetajet$ but with $\Gamma=\Gamma(\thetaobs)$ or $\Gamma=\Gamma(\thetajet)$ for on-axis and off-axis case, respectively.

For events with $\thetaobs > \thetajet$, we have also added data taken from the literature for a few well-known outliers of the AR, namely 
GRB980425 \citep[z=0.008,][]{amati06}, 
GRB 171205A/SN2017iuk \citep[z=0.037,][]{delia18}, 
GRB061021   \citep[z=0.3463,][]{nava2012},
GRB031203 \citep[z=0.106,][]{martone17}, and GRB080517 \citep[z=0.09,][]{stanway15}.

We find that their position in the
$\epi-\eiso$ plane is more consistent with the 
theoretical one derived for off-axis sources, and for which
the slope is $\epi \propto \eiso^{0.25}$.
This result strengthens the claim  that the outliers  of the AR are likely not intrinsically sub-luminous GRBs, but simply off-axis events which could be detected because of their nearness \citep{ruiz05, ghisellini06}.
The simple geometric argument has additionally the advantage of avoiding to search for other unknown  physical properties at the origin of the observable quantities.
Our simulations endorse the interpretation that the 
AR relation arises from the observation  of on-axis, highly-relativistic jets and originates from relativistic kinematics effects of sources with given
distribution of the Lorentz $\Gamma$-factor, the latter  playing the role
of leading parameter.
On the other hand, the AR observed dispersion is due to intrinsic dispersion 
of GRB properties such as the comoving-frame spectral emissivity shape and the typical radius $\rjet$ where the bulk of observed radiation during the prompt phase is released.
For off-axis events, a correlation in the $\epi-\eiso$ 
plane is still expected, but with a different slope $(\sim 0.25)$
whose value is  closer to 1/3 such as for the cannonball model  
under the same off-axis assumption \citep{dado2019}.

It is also worth noticing that for both on-axis and
off-axis sources, the assumption of a gaussian distribution
of the $\Gamma$-factors (if $E_0$ has a narrow distribution as well)
leads to a clustering of points in the top-right and
bottom-left part of the $\epi-\eiso$ diagram, respectively.
This is actually observed in the true data \citep[e.g.,][]{amati08}
and we claim that this is not due to observational bias effects, but arises
from the intrinsic properties of the GRB population.

\section{Discussion}
\label{discussion}

Despite the huge amount of theoretical work done until now to describe
the spectral emission of the GRB prompt phase, the scientific community
still uses phenomenological models for the X-ray spectral fitting.
As mentioned earlier, \grbcomp\ was the first physical model released
for the \xspec\ package.
The model is based on hydro-dynamical simulations performed by
\cite{chardonnet2010} who investigated the SN formation
due to pair-instability in very massive stars ($M\ga 200 M_{\odot})$,
a phenomenon which has received observational evidence \citep{gal-yam2009a, gal-yam2009b}.

The bulk of the emission in \grbcomp\ is due to Comptonization of blackbody-like seed photons ($\ktbb \sim$ few keV)  by  a Maxwellian population of hot electrons ($\kte \sim 100$ keV) moving outward the stellar
surface at sub-relativistic speed.
However, an association between pair-instability SN
and GRBs still lacks, and \grbcomp, albeit successful
in fitting data and providing results consistent
with simulations, appears strongly dependent on the
GRB progenitor class, besides the fact of working
out of the relativistic paradigm.

To overcome the phenomenological approach in the spectral analysis and work within the well-consolidated relativistic framework, we developed a numerical model which assumes  emission from a top-hat jet using the single-pulse approximation \citep{yamazaki03}.
Note that if the relativistic outflow is viewed on-axis 
($\thetaobs < \thetajet$), the observer can see only the top-hat surface and  there is essentially no  difference between a cone-like geometry extended over the radial distance from the center of the system, and a geometrically thin shell.

We now briefly discuss the reliability of the best-fit parameters while using the model for X-ray spectral analysis.
First let us define $F_{\rm CR} (\eobs, \Vec{P_0}, t)$ as the observed single-pulse light curve
due to curvature radiation at some energy $\eobs$, and
for a given set of parameters $\Vec{P_0}$.
It is possible to reformulate equation (\ref{angular_integral_flux}) as

\begin{equation}
 F_{\rm CR}(\eobs,\Vec{P_0}) =   \frac{ 1 }{ \tspa}  \int^{\tspa}_{0} F_{\rm CR} (\eobs, \Vec{P_0},t) dt,
 \label{flux_tspa}
\end{equation}

where $\tspa$ is the single-pulse duration in the observer frame defined in equations (\ref{tspa_on}) and (\ref{tspa_off}).
\noindent
On the other hand, the time-average spectrum over a general observed interval $\tgrb$ can be written as

\begin{equation}
 F(\eobs) =   \frac{ 1 }{ \tgrb}  \int^{\tgrb}_{0} F[\eobs, \Vec{P}(t),t] dt,
 \label{flux_true_data}
\end{equation}

where now $F[\eobs, \Vec{P}(t),t]$ is the true source light
curve and $\Vec{P}(t)$ is the time-dependent array of
parameters describing the spectrum.
The SPA model best-fit parameters are thus the ones which minimize the
difference between the right-hand terms in equations 
(\ref{flux_tspa}) and (\ref{flux_true_data}).
In this context, the array $\Vec{P_0}$ is a proxy of
$<\Vec{P}(t)>$, where the latter quantity is to be intended
as averaged over  $\tgrb$.

In a subsequent paper, we will present detailed time-dependent
results in the framework of SPA for on-axis and off-axis
events, together with mathematical tools for reproducing 
 light curves as compliant as possible with observations.


Despite the above described approximation, we outline how
the proposed model is able to naturally reproduce the observed $\epi-\eiso$ relation (AR) for on-axis events, at the same time providing a straightforward explanation for the outliers in terms
of simple viewing angle effects. 
This results strengthen the idea that the main 
characteristics of the sources are caught 
from the observational point of view, 
allowing to consider with good confidence the physical and/or
geometrical  parameters inferred from the spectral fitting procedure.

For pratical purposes, it is also important to point out that for $\thetaobs < \thetajet$ the observed peak energy $\ep$ and the flux 
are essentially independent on $\thetaobs$ (see Figure \ref{epeiso_vs_thobs}), and this
allows to reduce the space parameter dimension
by one degree of freedom by setting $\thetaobs$ to any value from 0 to $\thetajet$ in the X-ray spectral
fitting procedure for all GRBs obeying the AR, i.e. are seen on-axis.
The latest condition can be preliminarly tested by performing spectral fitting with e.g., the usual Band function
\citep{band93} and checking the position of the GRB in the $\epi-\eiso$ plane.
Moreover, under the same on-axis condition, most of the contribution to the flux comes from a region of width $\theta \sim 1/\Gamma$
centered along the direction to the observer 
(see Figure \ref{surface_brightness}), which is independent of the jet opening angle as long as $\thetajet \ga 1/\Gamma$.
This allows to further keep $\thetajet$ frozen to reasonable values
(let say $10^{\circ}-20^{\circ}$) during the fit, further lowering the number of free parameters.
Some degree of degeneracy is instead expected between the comoving-frame break energy $E_0$ (see equation \ref{sbpl_model}) and the $\Gamma$-factor as 
$\ep \propto E_0 \Gamma$ and $\liso \propto E_0 \Gamma^2$.
In this case, one should try to leave free both parameters and evaluate the magnitude of errors of the best-fit values or keep free one of the two quantities
from other independent evaluations.

Another point which deserves to be outlined is that the comoving-frame emissivity and the jet $\Gamma$-factor are here treated as
potentially independent parameters. 
Actually, for the internal shock model particle acceleration as well as magnetic field values of the emission zone depend on the hydrodynamical conditions
of the shocks forming between colliding shells. These in turn depend on the relative shell velocities and densities \citep[e.g.,][]{daigne1998, bosnjak2009}. The spectral shape and
normalization in the comoving frame are related to
the final $\Gamma$-factor of each couple of merged shells. 
It is however very difficult to provide within the context of a  model for spectral fitting some analytical or numerical dependencies of the comoving-frame emissivity parameters on the $\Gamma$-factor. This would indeed require a different  approach  to the problem, with a set of coupled radiative transfer and hydrodynamical simulations from which eventually  deriving explicit correlations to be
tabulated and later imported into a model which, we outline again, needs
to achieve a trade-off between computational speed and complexity.
At the observational level, correlations between the jet $\Gamma$-factor and local emissivity need to be derived downstream from
the model best-fit parameters.

\noindent 

The presented model, albeit focused on the GRBs, can be
considered \emph{general}, having no limitations in the 
relativistic outflow $\Gamma$-values. A possible drawback is given by the fact that the emission from the lateral
walls of a jet is neglected; this assumption is expected to essentially have little or no effect for $\thetaobs \leq \thetajet$, while for
off-axis events (if detectable) the fluxes
computed from the model should provide a lower limit
to the actual values, in particular for jets surrounded by 
a sub-relativistic cocoon \citep{kathirgamaraju18}. 
This would require however calculations of the emissivity
profile across the jet radial direction, with definition
of a  $\Gamma(R)$-law, which is outside the scope of the present work.
Note however that emission from the lateral walls may be important  for a jet having 
an appreciable radial extension, while for geometrical thin configurations  with $\Delta R/R <<1$ (i.e. a shell) the top-hat emission  here adopted provides
a sufficiently good approximation.

\section{Conclusions}
\label{conclusions}
The main purpose of our work was to make available for the \xspec\ package the first relativistic non-phenomenological model for fitting
the spectra of GRBs during the prompt phase.
We thus developed a  model for reproducing the observed
spectra arising from the emission of a top-hat relativistic jet or a geometrically thin shell
using the single pulse approximation.
Despite unavoidable simplifications, necessary
to have reasonable computational times with the \xspec\ package, we have shown that the model reproduces the observed slope $\sim 0.5$ in the plane
$\epi-\liso$ or $\epi-\eiso$ (AR) for
on-axis events ($\thetaobs < \thetajet$), and this effect naturally arises from pure relativistic kinematic effects, no-matter on the emissivity law in the comoving frame, provided a peak energy is of course present in the EF(E) spectrum.
For off-axis events ($\thetaobs > \thetajet$) the slope
is instead $\sim 0.25$.
Many efforts have been made over years for explaining the
physical origin of the AR \citep[e.g.,][]{guida08, dermer2009, ghirlanda2012, titarchuk2012,vyas2020} as well as its outliers, and in this work the disentangling  between two distinct classes of observed events, which depends on the observer viewing angle, has been achieved  with a thorough mathematical and numerical treatment.

\noindent
We outline that the observational testing of this theoretical prediction can be also a very important scientific goal for the next generation of GRB observatories such as  THESEUS \citep{amati2018},  whose great enhanced sensitivity is expected to be  able
to catch a large sample of weak  off-axis with 
enough statistics to allow time-resolved spectral
analysis.

\section*{Acknowledgments}
This project has received funding from the European Union’s Horizon 2020 research and innovation program under the Marie Sklodowska-Curie grant agreement n. 664931 (RB). 

\section*{Data availability}
The source code of the model, labeled {\sc grbjet}, is available at the website address\\ https://heasarc.gsfc.nasa.gov/xanadu/xspec/newmodels.html

\bibliography{OGRB}
\bibliographystyle{mnras}

\end{document}